\documentclass[twocolumn]{aastex631}

\usepackage{dcolumn}
\usepackage{bm}
\usepackage{amsmath}
\usepackage{natbib}

\usepackage{aas_macros}
\usepackage{mathtools}
\usepackage{amsmath, amssymb}
\usepackage{breqn}
\usepackage[T1]{fontenc}
\usepackage{ae,aecompl}
\usepackage{tablefootnote}

\usepackage{graphicx}	
\usepackage{amsmath}	
\usepackage{amssymb}	
\usepackage{float}
\usepackage{multirow}
\usepackage{units}
\usepackage{placeins}
\usepackage{physics}
\usepackage{ulem}
\usepackage{ragged2e}
\usepackage{xcolor}
\begin{document}

\title{Influence of winds on shocked magnetized viscous accretion flows around rotating black holes}

\author{Camelia Jana}
\affiliation{Indian Institute of Technology Guwahati, Guwahati 781039, Assam, India}
\correspondingauthor{sbdas@iitg.ac.in}
\email{camelia\_jana@iitg.ac.in}

\author[0000-0003-4399-5047]{Santabrata Das}
\affiliation{Indian Institute of Technology Guwahati, Guwahati 781039, Assam, India}
\email{sbdas@iitg.ac.in}

\begin{abstract}
    We study global transonic solution for a relativistic, magnetized, viscous advective accretion flow around a rotating black hole, incorporating the effects of mass and angular momentum loss through winds. Our model considers dominant toroidal magnetic fields with synchrotron radiation as the primary cooling mechanism. To self-consistently model mass loss, the mass accretion rate is prescribed to decrease inward as a power-law with disk radius. With this, we solve the governing equations that describe the accretion flows in presence of winds and obtain the flow structure in terms of the inflow parameters (energy $\mathcal{E}$, angular momentum $\lambda$, plasma-$\beta$, accretion rate $\dot{m}$, and viscosity $\alpha_{\rm B}$), the wind parameters ($p$, governing mass loss; and $l$, governing angular momentum transport by winds), and the black hole spin ($a_{\rm k}$). Our analysis reveals that winds substantially modify the accretion flow leading to a significant decrease in disk luminosity. We specifically identify global solutions that admit standing shocks and find that winds profoundly alter shock properties, such as the shock radius ($x_{\rm s}$), compression ratio ($R$), and shock strength ($S$). Furthermore, we determine the critical wind parameter $p^{\rm crit}$ beyond which steady shock solutions cease to exist. We demonstrate that increased viscosity and strong angular momentum extraction by winds lead to reduce $p^{\rm crit}$. These findings evidently highlight a complex interplay between viscosity and winds in governing the dynamics of shock formation in accretion disks.    
\end{abstract}

\keywords{
\href{http://astrothesaurus.org/uat/14}{Accretion (14)}; \href{http://astrothesaurus.org/uat/1963}{Hydrodynamics (1963)}; \href{http://astrothesaurus.org/uat/159}{Black hole physics (159)}; \href{http://astrothesaurus.org/uat/2086}{Shocks (2086)}; 
} 

\section{Introduction}

Accretion is widely regarded as the fundamental mechanism powering the energetic cosmic phenomena observed in X-ray binaries (XRBs), active galactic nuclei (AGNs), and gamma-ray bursts (GRBs). Over the years, several theoretical models have been developed to explain these processes. In the standard thin accretion disk model proposed by \cite{Shakura-Sunyaev-1973}, the energy generated within the disk is efficiently radiated away, resulting in a geometrically thin and radiatively efficient disk. In contrast, the advection-dominated accretion flow (ADAF) model \citep[]{Narayan-Yi-1994, Narayan-Yi-1995} suggests that only a fraction of the dissipated energy escapes as radiation, while the remaining energy is advected inward with the accreting gas giving rise to a radiatively inefficient hot flow.

One of the most intriguing findings in accretion studies is the presence of winds and outflows, where a fraction of the inflowing matter is expelled from the disk rather than accreting onto the black hole, causing the mass accretion rate to vary across the entire radial extent of the disk.  Towards this, \cite{Blanford-Payne-1982} showed that winds can be launched along the magnetic field lines and efficiently extract angular momentum and energy from the inflowing material. Since winds remove mass, angular momentum, and energy from the disk, a comprehensive understanding of accretion dynamics in their presence is fundamentally important. Among the initial theoretical attempts, \cite{Chakrabarti-1999} estimated the mass outflow rate from an advective accretion disk, demonstrating that a fraction of the inflowing matter can be deflected at the centrifugal pressure-supported post-shock corona to form outflows. This study provided a physical framework establishing inflow–outflow coupling within the advective paradigm. Subsequently, \cite{Blandford-Begelman1999} proposed the advection-dominated inflow–outflow solution (ADIOS), representing a self-consistent theoretical model incorporating mass loss through winds, in which the accretion rate varies with radius as $\dot{M}_{\rm a} \propto x^{p}$, with $p$ ranging between $0$ and $1$. Numerical simulations later validated this power-law dependence of the mass accretion rate on radius \citep[]{Stone-etal1999,Ohsuga-etal-2005, Yuan-etal-2012a, Yuan-etal-2012b}, emphasizing that outflows are an intrinsic component of accretion dynamics around compact objects. \cite{Li-Cao-2009} demonstrated that magnetically driven winds induce a near power-law profile in the outer disk regions, while deviations occur near the inner disk, and that mass loss results in a reduction of the inflow temperature. Similarly, \cite{Bu-etal-2009} explored the effects of outflows in magnetized accretion flows, showing that the presence of outflows significantly modifies the accretion dynamics. Specifically, increasing both the angular momentum and energy of the outflow reduces the inflow temperature, while the rotational velocity of the accretion flow either increases or decreases depending on the strength of the magnetic field. \cite{Shadmehri-2008} and \cite{Abbassi-etal-2010} also reported that winds lead to lower inflow temperatures and faster rotation of the accretion flow. More recently, \cite{Tamilan-etal-2025} investigated self-similar MHD accretion solutions in the presence of winds and confirmed the power-law dependence of the mass accretion rate. In addition, the study showed that in the intermediate multicolour blackbody regime, spectral luminosity exhibits a distinct power-law dependence on frequency due to wind effects. Overall, all these studies indicate that winds play a crucial role in shaping accretion dynamics.

During the course of accretion, inflowing matter begins its journey with negligible velocity at the outer edge of the disk and ultimately approaches the event horizon with a velocity comparable to the speed of light. To satisfy this inner boundary condition, any accretion solution around a black hole must undergo a smooth transonic transition, where the flow changes its sonic state at one or more critical points. Depending on the outer boundary conditions, multiple critical points may exist, which are a necessary condition for the formation of shock waves. During the inward progression, the rotating matter experiences a centrifugal barrier that can induce a discontinuous transition in the flow properties, giving rise to a standing or oscillating shock provided the Rankine–Hugoniot conditions are satisfied \cite[]{Landau-Lifshitz-1959}. Because of the shock transition, the inflow slows down resulting in accumulation of matter in the vicinity of the black hole. In this region, a significant portion of the bulk kinetic energy is converted into thermal energy, giving rise to a hot and dense post-shock corona (PSC). This PSC serves as a natural Comptonizing medium, where soft photons from the pre-shock accretion disk undergo inverse Compton scattering with hot electrons, producing high energy radiation \cite[]{Chakrabarti-Titarchuk1995,nandi-etal-2012,Iyer-etal-2015,Nandi-etal-2018,Majumder-etal2023}. It is worth mentioning that shock-induced accretion solutions are thermodynamically preferred due to their high entropy content \cite[]{Becker-Kazanas-2001}. Investigations of accretion flows containing shocks have been carried out in both hydrodynamic \citep[]{Fukue-1987, Chakrabarti-1989, Lu-etal-1999, Le-Becker-2005, Fukumara-Tsuruta-2004, Fukumura-Kazanas-2007, Dihingia-etal2018b, Dihingia-etal2019b, Sen-etal-2022, Singh-Das2024a, Singh-Das-2024b} and magnetohydrodynamic frameworks \citep[]{Koide-etal1999, Takahashi-etal2002, Takahashi-etal2006, Takahashi-Takahashi2010, Sarkar-Das2016, Sarkar-etal2018, Das-Sarkar-2018, Mitra-Das-2024, Singh-etal-2025}. However, the impact of winds on shock dynamics remains largely unexplored in astrophysical contexts.

Motivated by this, we investigate the effects of winds on relativistic, magnetized, viscous, dissipative, and advective accretion flows around a rotating black hole. We model the wind as a sink for both mass and angular momentum of the inflow, assuming that the mass accretion rate follows a power-law dependence on disk radius as a consequence of mass loss. The disk is assumed to be primarily threaded by a toroidal magnetic field, with synchrotron radiation serving as the dominant cooling mechanism. In addition, we adopt an effective potential \citep[]{Dihingia-etal2018a} that satisfactorily mimics the space–time geometry around a rotating black hole. Under these assumptions, we self-consistently solve the governing equations of the accretion flow in the presence of winds and obtain the global accretion solutions as functions of the energy ($\mathcal{E}$), angular momentum ($\lambda$), plasma-$\beta$ (which quantifies the magnetic activity within the disk), viscosity ($\alpha_{\rm B}$), mass accretion rate ($\dot{m}$), wind parameters $p$, and $l$ that regulates the angular momentum transport by the wind.

Our findings demonstrate that winds play a crucial role in governing accretion dynamics around black holes. In particular, the presence of winds leads to a noticeable reduction in disk luminosity. We also identify global accretion solutions harboring standing shocks and show that the wind parameters $p$ and $l$ strongly influence the shock properties, such as the shock location ($x_{\rm s}$), compression ratio ($R$), and shock strength ($S$). Since winds significantly alter the accretion structure, we further determine the upper limit of the wind parameter ($p^{\rm crit}$) based on the shock conditions, and find that $p^{\rm crit}$ strongly depends on both $l$ and $\alpha_{\rm B}$.

The paper is organized as follows. Section \ref{model} describes the underlying assumptions, governing equations, and the critical point analysis of the model. Section \ref{result} presents the results obtained from the analysis. Finally, Section~\ref{concl} provides a summary of the main findings and concluding remarks.

\section{Accretion flow model}
\label{model}

We consider an axisymmetric, advective accretion disk around a rotating black hole, assumed to lie on the equatorial plane in a steady state configuration. The system is described in a cylindrical polar coordinate framework ($x, \phi, z$), with the black hole resided at the origin. To facilitate the analysis, we adopt a natural unit system in which the gravitational constant ($G$), black hole mass ($M_{\rm BH}$), and speed of light ($c$) are each set to unity as $G=M_{\rm BH}=c=1$. In this framework, the radial coordinate, specific angular momentum, and energy are expressed in dimensionless units of $G M_{\rm BH}/c^2$, $G M_{\rm BH}/c$, and $c^2$, respectively.

\subsection{Governing equations for accretion in the presence of Wind}

We begin with a low angular momentum, relativistic, viscous, magnetized, dissipative, accretion flow around a BH in the presence of wind. To incorporate the magnetic field structure within the disk, we assume it to be turbulent in nature and predominantly governed by its toroidal component \cite[]{Hirose-etal-2006, Machida-etal-2006}. Accordingly, the magnetic field inside the disk is represented as $\left< \boldsymbol{B} \right> = \left< B_{\phi} \right> \hat{\phi}$ \cite[]{Oda-etal-2007}, where `$\left< \hspace{1mm} \right>$' denotes azimuthal averaging, and $B_{\phi}$ refers to the azimuthal component of the magnetic field. With these considerations, the governing equations for the magnetized accretion flow, incorporating the effects of wind \cite[]{Faghei-Mollatayefeh-2012}, are given by,\\
\noindent (a) Continuity equation:
\begin{equation}
	\frac{1}{x} \frac{\partial}{\partial x} (x \rho v_x) +  \frac{\partial}{\partial z} ( \rho v_z) = 0,
	\label{cont1}
\end{equation}
where $\rho$ is the mass density, and $v_x$ and $v_z$ denote the radial and vertical components of the velocity, respectively. In this study, we assume that the inflow has no vertical motion ($v_z = 0$), and wind originates from just above the disk surface. Under this consideration, the integration of equation~(\ref{cont1}) in the vertical direction yields \cite[]{Xie-Yuan-2008},
\begin{equation}
\frac{\partial }{\partial x}(2 \pi \Sigma x v_x) + 4 \pi x \rho_{\rm w} v_{z, {\rm w}} = 0,
\label{cont2}
\end{equation}
where $\Sigma ~(= 2 \rho H)$ represents the vertically integrated surface density of the inflowing matter with $H$ being the local half-thickness of the disk, and the subscript `$\rm w$' denotes quantities associated with the wind. Following \cite{Riffert-Herold-1995,peitz-Appl-1997}, we estimate the disk height ($H$) as,
\begin{equation}
	H = \sqrt{\frac{ P x^3}{\rho \mathcal{F}}}; \quad \mathcal{F} = \frac{1}{(1 - \lambda \Omega)} \frac{ (x^2 + a_k^2)^2 + 2 \Delta  a_{\rm k}^2}{ (x^2 + a_{\rm k}^2)^2 - 2 \Delta  a_{\rm k}^2},
	\label{height}
\end{equation}
where $\Omega = [2 a_{\rm k} + \lambda (x - 2)]~[a^2_{\rm k}(x + 2) - 2 a_{\rm k} \lambda + x^3]^{-1}$ is the angular velocity, and $P$ denotes the total pressure. Subsequently, we define the sound speed of the inflow as $C_{\rm s}^2 = \gamma P / \rho$, where $\gamma$ is the adiabatic index of the flow. While a self‑consistent estimation of $\gamma$ based on the local thermodynamic state of the disk would be more realistic, we assume a constant adiabatic index for simplicity, adopting $\gamma = 4/3$ throughout the flow unless stated otherwise.

The mass loss rate into the wind is given by \cite[]{Knigge-1999},
\begin{equation}
	\dot{M}_{\rm w} (x)= \int_{x_{\rm h}}^{x} 4 \pi \,\dot{m}_{\rm w} \, x dx,
	\label{wind}
\end{equation}
where $\dot{m}_{\rm w}$ denotes the mass outflow rate per unit area, and $x_{\rm h}$ represents the distance just outside event horizon.  In equation~(\ref{wind}), the factor of $2$ accounts for mass loss from both sides of the disk. By combining equations~(\ref{cont2}) and~(\ref{wind}), we finally obtain \cite[]{Knigge-1999, Faghei-Mollatayefeh-2012},
\begin{equation}
	\frac{\partial \dot{M}_a}{\partial x} = 	\frac{\partial \dot{M}_{\rm w}}{\partial x},
	\label{cont3}
\end{equation}
where the mass accretion rate is defined as $\dot{M}_a = 2 \pi \Sigma x v$ with $v = - v_x$. Indeed, equation~(\ref{cont3}) indicates that, due to the presence of wind, the mass accretion rate is not constant but instead varies with radial distance. Therefore, we adopt a power-law dependence of the mass accretion rate on radial distance \cite[]{Blandford-Begelman1999}, which is given by,
\begin{equation}
	\dot{M}_a(x) = \dot{M}_a(x_{\rm ref}) \times \Big( \frac{x}{x_{\rm ref}}\Big)^p,
	\label{power2}
\end{equation}
where $p$ is the wind parameter. The reference radius $x_{\rm ref}$ lies in the range $x_{\rm h} < x_{\rm ref} \leq x_{\rm edge}$, where $x_{\rm edge}$ denotes the outer edge of the disk. Analytical considerations \cite[see ADIOS model of][]{Blandford-Begelman1999} suggest that, the wind parameter $p$ lies in the range $0 \leq p < 1$, such that the mass accretion rate decreases while the energy of the outflowing matter increases with decreasing radius. Numerical studies, however, suggest a typical range of $p \sim 0.4 - 0.8$ \cite[]{Yuan-etal-2012a, Yuan-Narayan-2014}. Considering these results, in this work, we treat $p$ as a free parameter varying between $0$ and $1$.

\noindent (b) Radial momentum equation:
\begin{equation}
	v\frac{dv}{dx} + \frac{1}{\rho}\frac{dP}{dx} + \frac{d \Psi_{\rm eff}}{dx} + \frac{\left \langle {B_{\phi}}^2 \right \rangle}{4\pi \rho x} = 0,
    \label{radialmom}
\end{equation}
where the total pressure $P$ is the sum of gas pressure ($P_{\rm gas}$) and magnetic pressure ($P_{\rm mag}$), such that $P = P_{\rm gas} + P_{\rm mag}$. The gas pressure is given by $P_{\rm gas} = R \rho T / \mu$, where $R$, $T$, and $\mu$ denote the universal gas constant, the local flow temperature, and the mean molecular weight, respectively. In this study, we adopt $\mu = 0.5$, appropriate for a fully ionized plasma. The magnetic pressure is calculated as $P_{\rm mag} = \frac{\left\langle B_{\phi}^2 \right\rangle}{8\pi}$. Furthermore, we define the plasma-$\beta$ parameter as the ratio of gas pressure to magnetic pressure, $\beta = P_{\rm gas} / P_{\rm mag}$, which allows the total pressure to be expressed as $P = P_{\rm gas}(\beta + 1)/\beta$. In equation \eqref{radialmom}, we adopt the effective potential \cite[$\Psi_{\rm eff}$,][]{Dihingia-etal2018a}, which satisfactorily approximates the space-time geometry around a rotating BH. Its form on the equatorial plane is given by,
\begin{equation}
	\Psi_{\rm eff} = \frac{1}{2}{\rm ln}\left[\frac{x \Delta}{a_k^2(x +2) - 4a_{k} \lambda + x^{3} - \lambda^{2}(x-2)}\right],
\end{equation}
where $\lambda$ denotes the specific angular momentum (hereafter referred to as angular momentum), and $\Delta = x^2 - 2x + a_{\rm k}^2$, with $a_{\rm k}$ representing the Kerr parameter that characterizes the spin of the BH. It is worth mentioning that we approximate the spacetime around a rotating black hole by using an effective potential in the Newtonian limit in the radial momentum equation, instead of the trans‑relativistic formulation in which the potential is more accurate \cite[]{Dihingia-etal2018a}.

\noindent (c) Angular momentum conservation equation:
\begin{align}
&	\frac{\partial}{\partial x}\Big[   x \Sigma v \lambda  + x^2 T_{x \phi}\Big] - \frac{l \lambda}{2 \pi} \frac{\partial \dot{M}_{\rm w}}{\partial x} = 0,
	\label{angular}
\end{align}
where $T_{x \phi}$ denotes the vertically integrated viscous stress associated with the transport of angular momentum. In this analysis, we consider the total stress to be predominantly dominated by the $x\phi$ component. Moreover, following the simulation results of \cite{Machida-etal-2006}, in an advective accretion flows with significant radial motion, $T_{x \phi}$ can be expressed as 
\begin{equation}
	T_{x\phi} = \frac{<B_{x}B_{\phi}>}{4\pi}H = -\alpha_{\rm B}(W + \Sigma v^2),
	\label{viscosity}
\end{equation}
where $W ~(= 2  P H)$ denotes the vertically integrated total pressure \cite[]{Matsumoto-etal-1984} and $\alpha_{\rm B}$ is the proportionality constant. Following the work of \cite{Shakura-Sunyaev-1973}, we refer $\alpha_{\rm B}$ as the global viscosity parameter. The third term of equation (\ref{angular}) corresponds to the sink of angular momentum due to the wind, where `$l$' is defined as the ratio of the angular momentum of the outflowing matter to that of inflowing matter as \cite[]{Knigge-1999, Xie-Yuan-2008, saedi-Faghei-2022},
\begin{equation}
	\lambda_{\rm w}(x) = l \lambda (x).
\end{equation}
The parameter $l$ quantifies the extent of angular momentum transported away by the wind matter. The value of $l$ is not well constrained and is expected to depend on the underlying accretion–ejection dynamics. Self-similar solutions of advection-dominated accretion flows (ADAFs) suggest that outflows generally possess a lower specific angular momentum than the inflowing matter \citep[]{Narayan-Yi-1995b,Narayan-Yi-1995, Xu-Chen1997, Blandford-Begelman2004}, and this finding is further corroborated by numerical simulations \citep[]{Stone-etal1999, Chatterjee-Narayan2022}. In contrast, two-dimensional inflow–outflow models \citep[]{Zahra-etal2020} reveal that the mass-flux weighted angular momentum of the inflow can in fact be lower than that of the outflow suggesting that winds extract angular momentum from the disk. Furthermore, magnetic fields can couple the inflowing and outflowing plasma and enhance the angular momentum transport from the accretion flow to the ejected material \citep[]{Stone-Pringle2001, Blandford-Begelman2004}. Owing to these competing physical mechanisms and theoretical uncertainties, we parameterize $l$ within the range $0.8 \leq l \leq 1.2$ for this analysis, unless stated otherwise.
 
\noindent  (d) Entropy generation equation:
\begin{equation}
	\Sigma v T \frac {ds}{dx}=\frac{\Sigma v}{\gamma-1}
	\left(\frac{1}{\rho}\frac{dP_{\rm gas}}{dx} -\frac{\gamma P_{\rm gas}}{\rho^2}\frac{d\rho}{dx}\right)=Q^{-}  - Q^{+},
	\label{entropy}
\end{equation}
where $s$ denotes the specific entropy, and $Q^{+}$ and $Q^{-}$ represent the heating and cooling rates, respectively. Numerical simulations \cite[]{Hirose-etal-2006, Machida-etal-2006} further indicate that the accretion flow can be heated by the thermalization of magnetic energy through the process of magnetic reconnection, which can be expressed as \cite[]{Sarkar-etal2018,Dihingia-etal-2020, Jana-Das-2024}.
\begin{equation}
	Q^{+} = \frac{<B_{x}B_{\phi}>}{4\pi} x H \frac{d\Omega}{dx} = 
	-\alpha _{\rm B}(W + \Sigma v^2) x \frac{d\Omega}{dx}.
	\label{heatgain}
\end{equation}

In contrast, the accretion flow is cooled through several radiative processes, including bremsstrahlung, synchrotron, and Comptonization. Among these, bremsstrahlung is generally regarded as an inefficient coolant \cite[]{Chattopadhyay-Chakrabarti-2000}, while Comptonization requires a two-temperature plasma description, which is beyond the scope of this study. However, we note that Comptonization is likely become important in the inner disk regions. Since our investigation focuses on magnetized accretion disks, synchrotron radiation emerges as the predominant cooling process, and the corresponding cooling rate is given by \cite[]{Shapiro-Teukolsky-1983},
\begin{equation}
	Q^{-} = \frac{ S' a^5  \rho  H }{ v} \sqrt{\frac{\mathcal{F}}{ x^5}}\frac{\beta^2}{(1+\beta)^3},
	\label{heatcool}
\end{equation}
with 
$$
S' = 1.4827 \times 10^{18} \dot{m}_{\rm ref} \left( \frac{x}{x_{\rm ref}} \right)^p \frac{ \mu^2 e^4}{ m_{e}^3\gamma^{5/2}}\frac{1}{\ GM_{\odot}c^3},
$$
where $m_{\rm e}$ and $e$ represent the mass and charge of electron, respectively. The accretion rate ($\dot m$) is scaled with the Eddington rate ($\dot{M}_{\rm Edd}$) such that ${\dot{m}_{\rm ref} = \dot{M}_{\rm a}(x_{\rm ref})/\dot{M}_{\rm Edd}}$, where $\dot{M}_{\rm Edd} = 1.39 \times 10^{17}\, (M_{\rm BH}/M_{\odot})$ g s$^{-1}$, and $x_{\rm ref}$ stands for the reference radius. Here, we assume strong electron–ion coupling, resulting in a single temperature accretion flow. In realistic scenario, however, electrons being much lighter than ions are expected to attain lower temperatures, particularly in the vicinity of the event horizon \citep[]{Dihingia-etal2018b, Dihingia-etal-2020, Sarkar-Chattopadhyay-2022}. To account for this effect in a simplified manner, we estimate the electron temperature from the single temperature solution using $T_{\rm e} = \sqrt{m_{\rm e}/m_{\rm p}}\,T_{\rm p}$ \citep[]{Chattopadhyay-Chakrabarti-2000}, and adopt this prescription throughout the disk. Here, $T_{\rm p} \,(= T)$ and $m_{\rm p}$ denote the proton temperature and mass, respectively. In this study, we further assume that the wind is primarily powered by an external driving mechanism, rather than by the internal dissipation of accretion energy. Accordingly, energy loss via the wind is systematically excluded from our analysis.

\noindent (e)  Radial advection equation of the toroidal magnetic flux:

\begin{equation}
	\frac {\partial \left<B_{\phi}\right>\hat{\phi}}{\partial t} = {\bf \nabla} \times
	\left({\vec{v}} \times \left<B_{\phi}\right>\hat{\phi} -{\frac{4\pi}{c}}\eta {\vec{j}}\right),
	\label{induction}
\end{equation}
where $\vec{v}$, $\vec{j}$, and $\eta$ denote the velocity vector, the current density, and the resistivity of the flow, respectively. Because of the large length scale, the Reynolds number of an accretion disk is typically very high, allowing us to neglect the magnetic diffusion term. For simplicity, we also neglect the dynamo term and assume that the azimuthally averaged magnetic field vanishes at the disk surface. Under these assumptions, the advection rate of the magnetic flux is obtained as \cite[]{Oda-etal-2007},
\begin{equation}
	\dot{\Phi} = - \sqrt{4\pi}v H {B}_{0} (x),
	\label{magflux}
\end{equation}
where $B_{0}(x)$ represents the azimuthally averaged magnetic field confined to the disk equatorial plane. In a realistic scenario, $\dot{\Phi}$ is not a conserved quantity because of the dynamo and magnetic diffusion terms. However, incorporation of these terms is highly complex and lies beyond the scope of this paper. Meanwhile, global 3D MHD simulations demonstrate that $\dot{\Phi}$ inversely varies with radius as $\dot{\Phi} \propto 1/x$, when the disk is in quasi-steady state \cite[]{Machida-etal-2006}. To capture this behavior, we adopt the following parameterization as \citep[]{Oda-etal-2007},
    \begin{equation}
        \dot{\Phi}(x) =  \dot{\Phi}_{\rm edge} \Big(\frac{x}{x_{\rm edge}}\Big)^{-\zeta},
    \end{equation}
where $\dot{\Phi}_{\rm edge}$ denotes the advection rate of toroidal flux at $x_{\rm edge}$. In this study, we choose $\zeta = 1$ all throughout unless stated otherwise.

\subsection{Critical Point Analysis}
\label{critical}

The journey of accreting matter begins with an almost negligible velocity from the outer edge of the disk. As the flow spirals inward, it is eventually plunged through the event horizon at the speed of light. This inner boundary condition, imposed by the very nature of the event horizon, requires that any physically consistent accretion solution around a black hole must undergo a smooth sonic state transition at least once, if not multiple times, along its trajectory. To obtain global transonic accretion solution, we follow the methodology introduced by \cite{Chakrabarti-1989,Das-etal-2001a}, and using equations (\ref{power2}), (\ref{radialmom}), (\ref{angular}), (\ref{entropy}), and (\ref{magflux}), we compute the gradients of flow variables, expressed as,
\begin{align}
& \frac{dv}{dx}  = \frac{N (x, v, C_{\rm s}, \lambda, \beta)}{D (x, v, C_{\rm s}, \lambda, \beta)},
\label{grad_v}\\
& \frac{d C_{\rm s}}{dx}  = C_{\rm s 0} + C_{{\rm s} v} \frac{dv}{dx},
\label{grad_a}\\
& \frac{d \lambda}{dx}  = \lambda_0 + \lambda_{ v} \frac{dv}{dx}, 
\label{grad_l}\\
& \frac{d \beta}{dx}  = \beta_0 + \beta_{ v} \frac{dv}{dx}.
\label{grad_b}
\end{align}
In equations (\ref{grad_v}-\ref{grad_b}), the explicit expressions of the numerator $N(x, v, C_{\rm s}, \lambda, \beta)$, the denominator $D(x, v, C_{\rm s}, \lambda, \beta)$, and the coefficients $C_{\rm s0}$, $C_{{\rm s}v}$, $\lambda_0$, $\lambda_v$, $\beta_0$, and $\beta_v$ are provided in the Appendix. At the critical point ($x_{\rm c}$), the numerator $N$ and the denominator $D$ of equations (\ref{grad_v}) vanish simultaneously and the velocity gradient takes the indeterminate form as $\frac{dv}{dx} = 0/0$. However, for accretion solutions to remain physically viable, $\frac{dv}{dx}$ must be real and finite along the streamline of the flow. Accordingly, we apply L’H$\hat{o}$pital’s rule to calculate $\left( \frac{dv}{dx} \right)_{x_c}$ at $x_{\rm c}$. For a chosen set of model parameters, we obtain two values of $\left( \frac{dv}{dx} \right)_{x_c}$. Based on these values, the nature of the critical points is then classified as saddle type, nodal type and spiral type \cite[]{Das-etal-2001a}. It is worth mentioning that only saddle-type critical points are physically relevant \cite[]{Kato-etal-1993, Das-2007}, with the negative values of $\frac{dv}{dx}$ results in the accretion solution. Furthermore, when the critical point is located close to the black hole, it is referred to as the inner critical point ($x_{\rm in}$), while one situated farther away is termed the outer critical point ($x_{\rm out}$).

\section{Results}
\label{result}

Given the pivotal role of mass loss in shaping accretion solutions, it is imperative to investigate how the presence of wind alters the physical characteristics of the inflowing material. In this section, we present our analysis of black hole accretion dynamics in the presence of mass outflows, with the primary aim of assessing the impact of wind on the overall flow properties. To this end, we explore two distinct regimes such as (a) $l < 1$ and (b) $l > 1$ and examine the corresponding global solutions for a range of wind parameters ($p$).

\subsection{Global Solution tropology}
\label{solution}

\begin{figure}
    \begin{center}
        \includegraphics[width=\columnwidth]{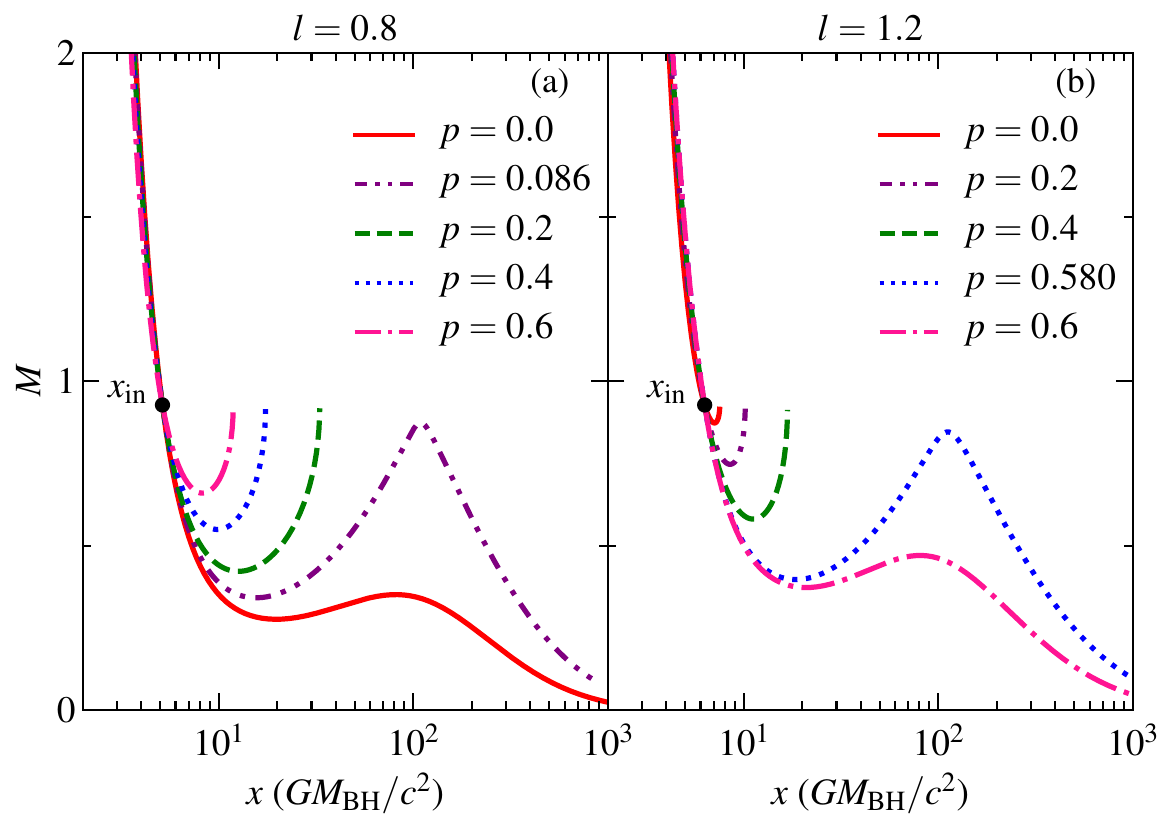}
    \end{center}
    \caption{Variation of Mach number ($M$) with radial distance ($x$) for different wind parameters ($p$) for flows passing through the inner critical points ($x_{\rm in}$). Here, we choose the model parameters as $\alpha_{\rm B}=0.02$, $a_{\rm k}=0.0$, $\beta_{\rm in}=100.0$, and $\dot{m}_{\rm in}=0.01$. The left panel corresponds to $l=0.8$, $x_{\rm in}=5.14$, and $\lambda_{\rm in}=3.42$, while the right panel represent the case with $l=1.2$, $x_{\rm in}=6.30$, and $\lambda_{\rm in}=3.46$. See text for the details.
    }
    \label{fig01}
\end{figure}

To obtain the global solution, we numerically solve the coupled differential equations (\ref{grad_v}–\ref{grad_b}) using a set of input parameters. Here, $a_{\rm k}$, $\alpha_{\rm B}$, $l$, and $p$ are treated as global parameters, whereas $\mathcal{E}$, $\lambda$, $\beta$, and $\dot{m}$ are specified locally at a chosen reference radius ($x_{\rm ref}$). Since any physically viable accretion flow onto a black hole must pass through a critical point ($x_{\rm c}$), we use this location as the reference point for our calculation. At $x_{\rm c}$, the flow variables ($v_c$ and $C_{\rm sc}$) are determined via critical point analysis, as described in \S\ref{critical}. With these critical conditions and model parameters, we integrate the governing equations outward to the disk's outer boundary ($x_{\rm edge}$) and inward toward the event horizon ($x_{\rm h}$). These two branches are then seamlessly connected to yield the complete, transonic global solution.

\subsubsection{Accretion solution through inner critical point {\rm ($x_{\rm in}$)}}
\label{sol-xin}

In Fig. \ref{fig01}, we display the global transonic accretion solution obtained by choosing the inner critical point ($x_{\rm in}$) as the reference point. The figure delineates the radial profile of the Mach number ($M$) as a function of logarithmic distance, highlighting the dynamical behavior of the accreting matter around a black hole. In panel (a), we fix the inner critical point at $x_{\rm in} = 5.14$ with $\lambda_{\rm in} = 3.42$, $\dot{m}_{\rm in} = 0.01$ and $\beta_{\rm in} = 100$, and set $l = 0.8$, $\alpha_{\rm B} = 0.02$ and $a_{\rm k} = 0.0$. The solid (red), dot-dot-dashed (purple), dashed (green), dotted (blue), and dot-dashed (magenta) curves correspond to $p = 0.0,\ 0.086,\ 0.2,\ 0.4$, and $0.6$, respectively. In the absence of mass loss ($p = 0$), accretion solution connects the disk's outer edge ($x_{\rm edge}=1000$) and the horizon ($x_{\rm h}$), where the accretion flow starts with subsonic speed, accelerates inward, crosses the critical point, and smoothly transits to the supersonic regime before plunging into the black hole. As we gradually increase $p$, the nature of solution topology begins to change. It is worth noting that beyond a critical value $p = 0.086$, the solution can no longer extend to the outer edge, resulting in a closed solution, which on its own does not represent a viable global accretion flow. However, such closed solutions become physically meaningful when connected to another branch passing through the outer critical point ($x_{\rm out}$) via a shock transition (see \S \ref{shock_sec}). With further increase in $p$, the solution topology continues to shrink and eventually vanishes entirely. This analysis clearly demonstrates that for $l < 1$, increasing mass loss leads to the closure and eventual disappearance of the global accretion solution. In panel (b), we investigate the opposite regime by setting $l = 1.2$, $\alpha_{\rm B} = 0.02$, $a_{\rm k} = 0.0$, with $x_{\rm in} = 6.30$, $\lambda_{\rm in} = 3.46$, $\dot{m}_{\rm in} = 0.01$, and $\beta_{\rm in} = 100$. Interestingly, for $p = 0$ (solid red curve), the solution is closed and does not extend to the disk’s outer edge. However, as $p$ increases, the closed solution gradually expands, and eventually connects to the outer edge for $p = 0.580$ (dotted blue curve). Beyond this threshold, the solution continues to open up with further increase of $p$ dot-dash (magenta) curve, resulting in viable global accretion solutions. Thus, for $l > 1$, mass loss has the opposite effect that transforms a closed solution into a complete global accretion flow. It is important to note that the limiting value of $p$ separating closed and open topologies is not universal but depends on the choice of input parameters. Furthermore, we emphasize that the overall nature of the accretion solutions is strongly regulated by wind properties, particularly the parameters $p$ and $l$. Although we have used $a_{\rm k} = 0$ for this analysis, we stress that the qualitative nature of the results remains unchanged for spinning black holes.

\subsubsection{Accretion solutions through outer critical point {\rm ($x_{\rm out}$)} }
\label{sol-xout}

\begin{figure}
    \begin{center}
        \includegraphics[width=\columnwidth]{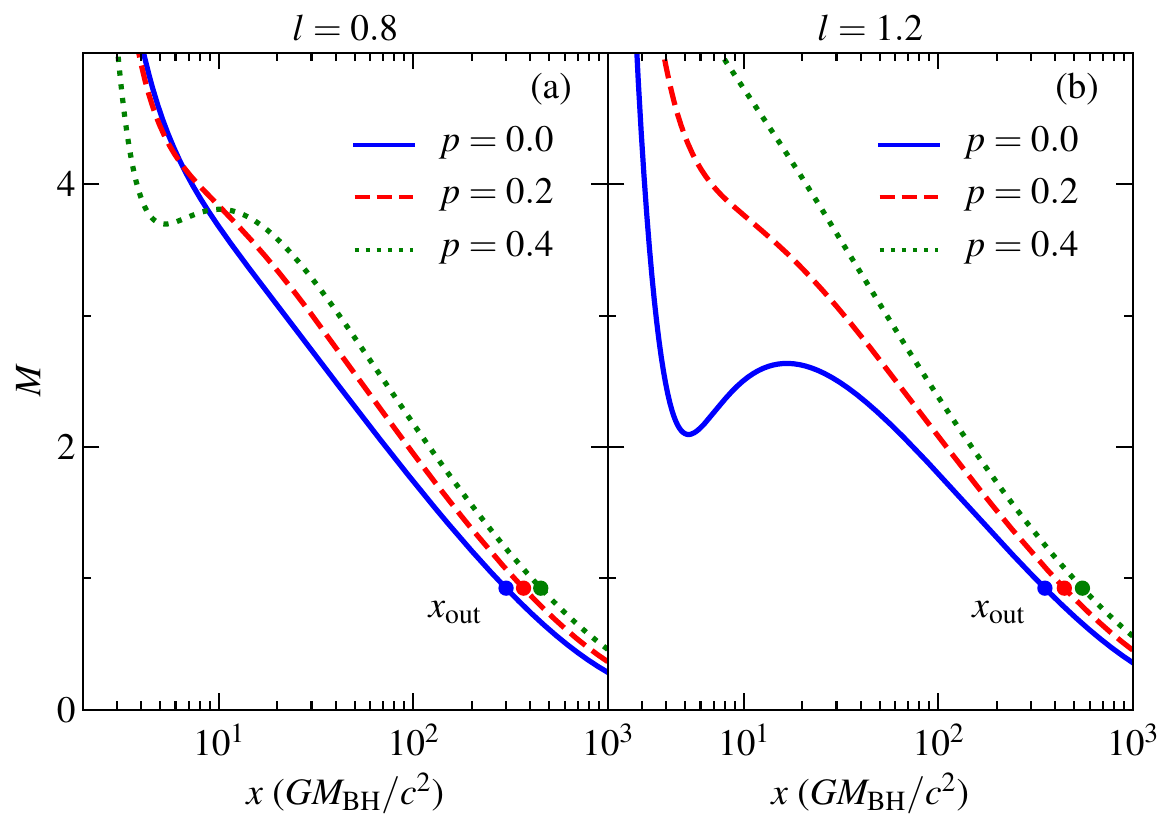}
    \end{center}
    \caption{Variation of Mach number ($M$) with radial distance ($x$) for different wind parameters ($p$) for flow containing outer critical points ($x_{\rm out}$). The flow is injected from $x_{\rm edge} = 10^3$ with $\dot{m}_{\rm edge} = 0.1$ and $\beta_{\rm edge} = 10^4$, while $\alpha_{\rm B} = 0.2$ and $a_{\rm k} = 0.0$ are kept fixed. In panel (a), we consider $l = 0.8$, $\mathcal{E}_{\rm edge} = 1.95 \times 10^{-3}$, and $\lambda_{\rm edge} = 3.60$, whereas in panel (b), $l = 1.2$, $\mathcal{E}_{\rm edge} = 1.58 \times 10^{-3}$, and $\lambda_{\rm edge} = 4.65$ are chosen. See text for the details. 
    }
    \label{fig02}
\end{figure}

Since the accreting matter commences its journey subsonically at large distances from the black hole, a physically consistent alternative approach is to fix the local flow parameters at the outer edge of the disk ($x_{\rm edge}$) and systematically vary the wind parameter ($p$) to assess its influence on the dynamical behavior of the flow. To accomplish this, we first specify the input parameters at the critical point ($x_{\rm c}$) and obtain the corresponding global transonic solution that extends from the black hole horizon to the outer edge ($x_{\rm edge}$), following the procedure described earlier. Here, we set $x_{\rm c} = x_{\rm out}$, as in this section we restrict our analysis to solutions passing through the outer critical point. Using the computed flow variables at $x_{\rm out}$, the local specific energy is evaluated as $\mathcal{E}(x) = \frac{v^2}{2} + \frac{C^2_{\rm s}}{\gamma - 1} + \Psi_{\rm eff} + \frac{\langle B_{\phi}^2 \rangle}{4\pi \rho}$. Alternatively, instead of initiating the solution from the critical point ($x_{\rm out}$), one may impose boundary conditions at the outer edge ($x_{\rm edge}$) and integrate inward by solving equations (\ref{grad_v})–(\ref{grad_b}), which yields an identical global solution. For illustration, we choose a fixed $x_{\rm edge}$ and adopt a black hole mass of $M_{\rm BH} = 10 M_{\odot}$ throughout this work unless stated otherwise, while exploring two representative cases, namely $l < 1$ and $l > 1$.

In Fig. \ref{fig02}, we illustrate the variation of the Mach number ($M$) for global accretion solutions as a function of radial distance ($x$), where panels (a) and (b) correspond to $l = 0.8$ and $l = 1.2$, respectively. Here, we fix the outer edge at $x_{\rm edge} = 1000$. In panel (a), we set $\alpha_{\rm B} = 0.02$, $ a_{\rm k}=0.0$, and the outer edge parameters as $\dot{m}_{\rm edge} = 0.1$, $\beta_{\rm edge} = 10^4$, $\mathcal{E}_{\rm edge} = 1.95 \times 10^{-3}$, and $\lambda_{\rm edge} = 3.60$. For $p = 0.0$, the global solution (solid blue) passes through the outer critical point at $x_{\rm out} = 300.07$, smoothly connecting the outer edge to the horizon. Increasing $p$ while keeping other parameters fixed, we obtain the dashed (red) and dotted (green) curves for $p = 0.2$ and $0.4$, with critical points at $x_{\rm out} = 369.27$ and $452.78$, respectively. In panel (b), we choose $\alpha_{\rm B} = 0.02$, $a_{\rm k} = 0.0$, and outer edge parameters $\dot{m}_{\rm edge} = 0.1$, $\beta_{\rm edge} = 10^4$, $\mathcal{E}_{\rm edge} = 1.58 \times 10^{-3}$, and $\lambda_{\rm edge} = 4.65$. The solid (blue), dashed (red), and dotted (green) curves correspond to $p = 0.0, 0.2$, and $0.4$, with the respective outer critical points at $x_{\rm out} = 353.57, 445.07$, and $551.50$. Taken together, these findings demonstrate that the global structure and dynamics of accretion solutions are strongly regulated by the wind parameter $p$ as similarly observed in Section \ref{sol-xin}.

\begin{figure}
    \begin{center}
        \includegraphics[width=\columnwidth]{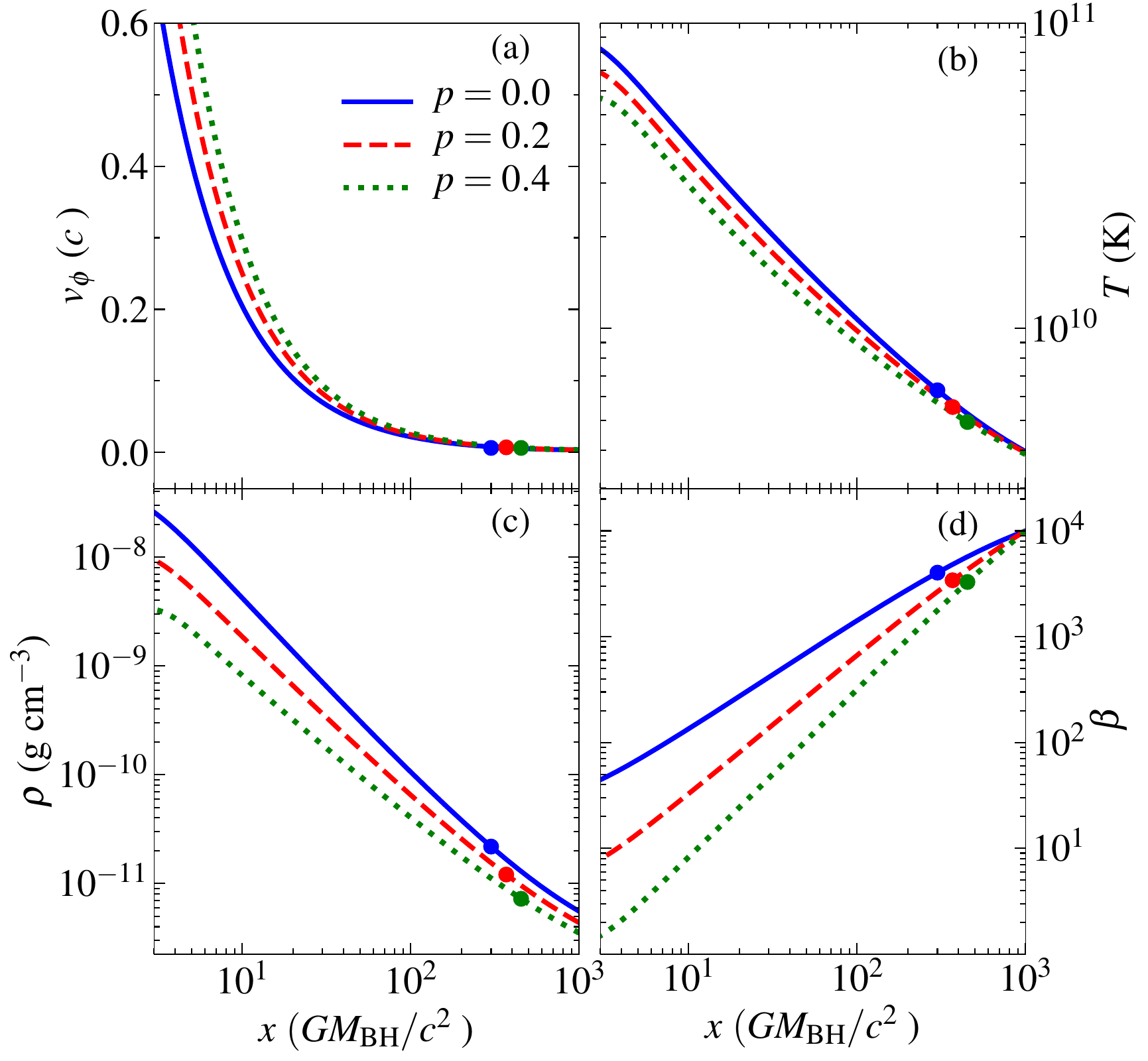}	
    \end{center}
    \caption{Radial variation of flow variables for different wind parameters $p$. Here, we fix $l=0.8$, $a_{\rm k}=0.0$, and $\alpha_{\rm B}=0.02$ and flow is injected from $x_{\rm edge}=1000$ with $\mathcal{E}_{\rm edge}=1.95\times10^{-3}$, $\lambda_{\rm edge}=3.60$, $\dot{m}_{\rm edge}=0.1$, and $\beta_{\rm edge}=10^4$. The solid (blue), dashed (red), and dotted (green) curves correspond to $p=0.0$, $0.2$, and $0.4$, respectively. See text for further details.
    }
    \label{fig03}
\end{figure}

In Fig. \ref{fig03}, we depict the radial profiles of key flow variables corresponding to the accretion solutions presented in Fig. \ref{fig02}a. In the figure, the solid (blue), dashed (red), and dotted (green) curves correspond to $p = 0$, $0.2$, and $0.4$, respectively, with filled circles indicating the location of the critical points. Fig. \ref{fig03}a illustrates the variation of the azimuthal velocity ($v_{\phi}$) as a function of radial distance ($x$). For $p = 0$, $v_{\phi}$ increases as the accretion flow approaches the black hole. With increasing $p$, this trend intensifies, leading to a higher $v_{\phi}$ throughout the disk. This behavior arises because, for $l < 1$, the outflowing material carries away less specific angular momentum than it initially possessed, effectively transferring the excess angular momentum back to the inflow. As a result, the accreting matter spins up, leading to enhanced rotational velocity in response to increased mass loss. During the accretion process, as the flow advects inward, it undergoes geometrical compression and gets heated up. In order to examine this, in Figs. \ref{fig03}b-c, we present the variation of temperature ($T$) and density ($\rho$) for the accretion solutions presented in Fig. \ref{fig02}a. We observe that both temperature ($T$) and density ($\rho$) increase as the flow approaches the event horizon. In Fig. \ref{fig03}b, we further notice that the temperature of the flow decreases with increasing mass loss rate. This evidently indicates that mass loss effectively cools down the disk. Similarly, in Fig. \ref{fig03}c, the density ($\rho$) of the flow decreases as the mass loss rate increases. This reduction is expected, as a fraction of the accreting material is diverted into the wind, thereby lowering the density of the remaining inflow. Finally, in Fig. \ref{fig03}d, we display the variation of the plasma-$\beta$ parameter, which decreases with increasing mass loss rate. This trend is resulted due to the combined effects of reduced temperature and density, which diminishes the gas pressure and, in turn, leads to intensely magnetized flows characterized by reduced plasma-$\beta$ values.
 
\begin{figure}
    \begin{center}
        \includegraphics[width=\columnwidth]{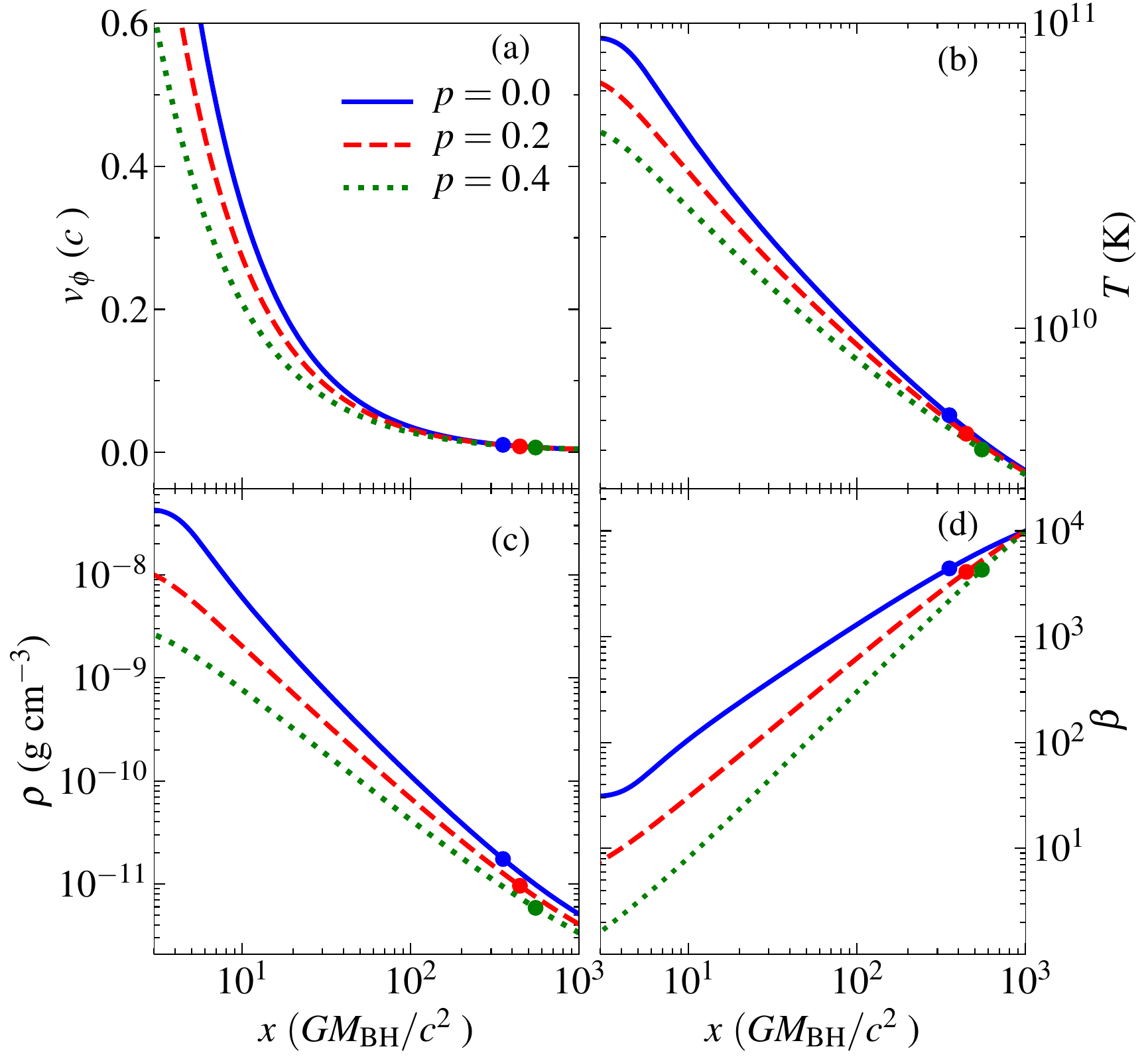}
    \end{center}
    \caption{Variation of flow properties with the radial distance ($x$) for wind parameters $p = 0.0$ (solid-blue), $0.2$ (dashed-red), and $0.4$ (dotted-green). Here, we choose $l = 1.2$ and the other input parameters are $\mathcal{E} _{\rm edge} = 1.58 \times 10^{-3}$, $\lambda_{\rm edge} = 4.65$, $\dot{m}_{\rm edge} = 0.1$, $\beta_{\rm edge} = 10^4$, $a_{\rm k} = 0.0$, and $\alpha_{\rm B} = 0.02$. See the text for the details.
    }
    \label{fig04}
\end{figure}

In Fig. \ref{fig04}, we examine how the wind influences the flow variables corresponding to the accretion solutions shown in Fig. \ref{fig02}b. The solid (blue), dashed (red), and dotted (green) curves correspond to results obtained for $p = 0$, $0.2$, and $0.4$, respectively. Fig. \ref{fig04}a illustrates the radial profile of $v_{\phi}$, which increases as flow accretes towards the BH, irrespective of the $p$ values. However, at any fixed radius, $v_{\phi}$ is reduced with increasing $p$. This trend arises because, for $l > 1$, the outflow efficiently extracts angular momentum from the accretion flow. Consequently, the accretion flow exhibits progressively slower rotation as the mass loss intensifies. In addition, we observe that the flow becomes cooler and more rarefied with higher $p$, as evident from the overall decrease in temperature ($T$) and density ($\rho$) depicted in Fig. \ref{fig04}b-c. Moreover, plasma-$\beta$ also decreases with increasing $p$ (see Fig. \ref{fig04}d), a consistent trend with that presented in Fig. \ref{fig03}. Altogether, these results clearly indicate that mass loss from the disk plays a regulatory role in governing the thermodynamic and magnetohydrodynamic structure of the accretion flow.

\begin{figure}
    \begin{center}
        \includegraphics[width = \columnwidth]{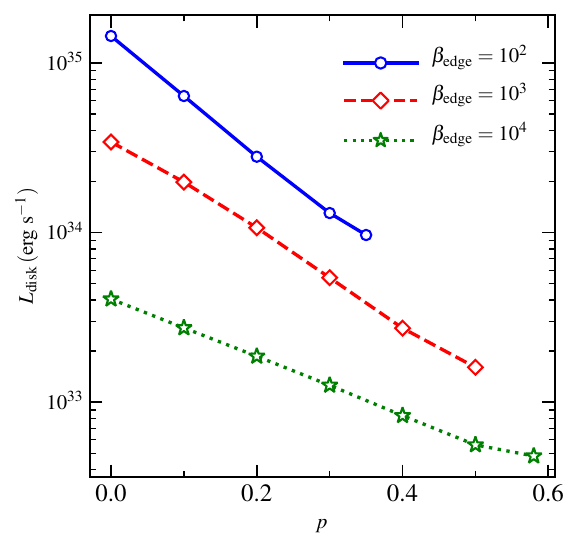}
    \end{center}
    \caption{Variation of the disk luminosity ($L_{\rm disk}$) as a function of wind parameter ($p$) for a set of plasma-$\beta$ parameters. Here, the flow is injected from $x_{\rm edge} = 10^3$ with $\mathcal{E}_{\rm edge} = 1.95 \times 10^{-3}$, $\lambda_{\rm edge} = 3.60$, $\dot{m}_{\rm edge} = 0.1$ and we set $l = 0.8$, $\alpha_{\rm B} = 0.02$ and $a_{\rm k} = 0.0$. Circles, diamonds, and asterisks connected with solid (blue), dashed (red), and dotted (green) lines correspond to $\beta_{\rm edge} = 10^2,$ $10^3$, and $10^4$, respectively. See text for the details.
    }
    \label{fig05}
\end{figure}

\subsection{Estimation of disk luminosity}

Therefore, mass loss in the form of winds is expected to significantly influence the radiative properties of the accreting matter, particularly the luminosity emerging from the disk. Motivated by this, in the subsequent analysis, we investigate how disk driven winds alter the luminosity of the accretion disk. Since synchrotron radiation is considered as the primary cooling mechanism in this study, the corresponding disk luminosity is expressed as,
\begin{equation}
    L_{\rm disk} =4 \, \pi \int^{x_{\rm edge}}_{x_{\rm h}} Q^{-} \, x \, dx.
    \label{bolo}
\end{equation}
Using equation (\ref{bolo}), we compute the disk luminosity by utilizing the accretion solutions, and the obtained results are presented in Fig. \ref{fig05}.

In the Fig. \ref{fig05}, we illustrate the variation of disk luminosity ($L_{\rm disk}$) with wind parameter ($p$) for different $\beta_{\rm edge}$. In doing so, we inject accreting matter from $x_{\rm edge}=1000$ with $\mathcal{E}_{\rm edge} = 1.95 \times 10^{-3}$, $\lambda_{\rm edge} = 3.60$, $\dot{m}_{\rm edge} = 0.1$, and fix $\alpha_{\rm B} = 0.02$, $l = 0.8$, and $a_{\rm k} = 0.0$. Here, the open circles, diamonds, and asterisks connected by solid (blue), dashed (red), and dotted (green) lines correspond to the results obtained for $\beta_{\rm edge} = 10^2$, $10^3$, and $10^4$, respectively. The results clearly indicate that for a given $\beta_{\rm edge}$, $L_{\rm disk}$ decreases as $p$ (equivalently mass loss) is increased. In reality, enhanced mass loss reduces the density of the accretion flow, which in turn lowers the radiative efficiency and suppresses the disk luminosity. The significant decrease in disk luminosity predicted by our wind driven model is consistent with observed anti-correlations in black hole binaries like GRS 1915+105, where major radio flares (indicating powerful ejections) are frequently followed by a sharp drop in X-ray flux, as the accretion energy is possibly channeled into the jet/wind \cite[]{Belloni-etal2000,Fender-etal-2004}. Moreover, for a fixed value of $p$, we find that $L_{\rm disk}$ remains higher for smaller plasma-$\beta$. This is expected, since a lower plasma-$\beta$ corresponds to stronger magnetic activity within the disk, thereby enhancing synchrotron emission and boosting disk luminosity. What is more is that beyond a critical limit of $p$, the accretion solution fails to connect the outer edge ($x_{\rm edge}$) to the event horizon ($x_{\rm h}$), resulting in closed accretion solutions passing through $x_{\rm out}$. Crucially, this threshold value of $p$ is not universal, rather it intricately depends on the input parameters chosen at the outer edge ($x_{\rm edge}$) of the accretion flow.

\begin{figure}
    \begin{center}
        \includegraphics[width = \columnwidth]{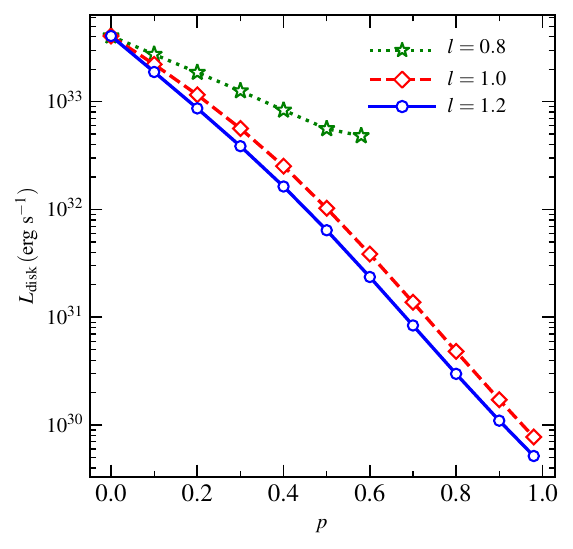}
    \end{center}
    \caption{Variation of the disk luminosity ($L_{\rm disk}$) as a function of wind parameter ($p$) for a set of $l$ parameters. Here, the flow is injected from $x_{\rm edge} = 10^3$ with $\mathcal{E}_{\rm edge} = 1.95 \times 10^{-3}$, $\lambda_{\rm edge} = 3.60$, $\dot{m}_{\rm edge} = 0.1$, $\beta_{\rm edge} = 10^4$, $\alpha_{\rm B} = 0.02$ and $a_{\rm k} = 0.0$. Open asterisks, diamonds, and circles connected with dotted (green), dashed (red), and solid (blue) lines correspond to $l = 0.8,$ $1.0$, and $1.2$, respectively. See text for the details. 
    }
    \label{fig06}
\end{figure}

Furthermore, we investigate how the removal of angular momentum by disk driven winds effects the radiative output of the accretion flow. To this end, we inject matter from $x_{\rm edge}=1000$ with $\mathcal{E}_{\rm edge} = 1.95 \times 10^{-3}$, $\lambda_{\rm edge} = 3.60$, $\dot{m}_{\rm edge} = 0.1$, $\beta_{\rm edge} = 10^4$, $\alpha_{\rm B} = 0.02$, and $a_{\rm k} = 0.0$, and explore the effect of mass loss ($p$) on the disk luminosity ($L_{\rm disk}$) for different values of $l$ parameters. The obtained results are displayed in Fig.~\ref{fig06}, where the open asterisks, diamonds, and circles connected with dotted (green), dashed (red), and solid (blue) lines correspond to the results obtained for $l = 0.8$, $1.0$, and $1.2$, respectively. The figure reveals that the disk luminosity diminishes progressively as winds carried away angular momentum from the accretion flow. Moreover, we observe that irrespective of $l \lessgtr 1$, the presence of winds invariably drives a monotonic suppression of luminosity with increasing wind strength ($p$). These finding suggests that the removal of angular momentum by winds plays a decisive role in regulating the radiative efficiency of accretion disks.

\subsection{Shock induced global solutions}
\label{shock_sec}

In Section 3.1, we point out that depending on the values of $p$, accretion solutions passing through the inner critical points often fail to extend up to the outer edge of the disk. However, such closed solutions can connect with other accretion branches passing through the outer critical points via a shock transition, provided the Rankine–Hugoniot conditions (RHCs) are satisfied \cite[]{Landau-Lifshitz-1959}. In rotating accretion flows, the centrifugal barrier naturally opposes gravity and the supersonic inflow is slowed down causing the accumulation of matter in the vicinity of the black hole. However, matter accumulation cannot be sustained indefinitely. Once a threshold is reached, the interplay between the opposing forces under favorable conditions (RHCs) in supersonic flow triggers a discontinuity in the flow variables along the subsonic branch, which is identified as a shock transition. After the shock, flow gradually gains its velocity and ultimately plunges into the black hole supersonically after crossing the inner critical point. It is worth mentioning that the shock-induced global accretion solutions are thermodynamically favored, as they possess higher entropy compared to their shock-free solution \cite[]{Becker-Kazanas-2001}. The RHCs are expressed through the conservation of energy flux, mass flux, momentum flux, and magnetic flux across the shock front. In this work, we assume the shock to be infinitesimally thin, and therefore, we impose the following conditions: (a) $\mathcal{E}_{+} = \mathcal{E}_{-}$, (b) $\dot{M}_{a+} = \dot{M}_{a-}$, (c) $W_{+} + \Sigma_{+} v_{+}^{2} = W_{-} + \Sigma_{-} v_{-}^{2}$, and (d) $\dot{\Phi}_{+} = \dot{\Phi}_{-}$, where the subscripts `$-$' and `$+$' denote the upstream and downstream quantities across the shock front, respectively.

\begin{figure}
    \begin{center}
        \includegraphics[width=\columnwidth]{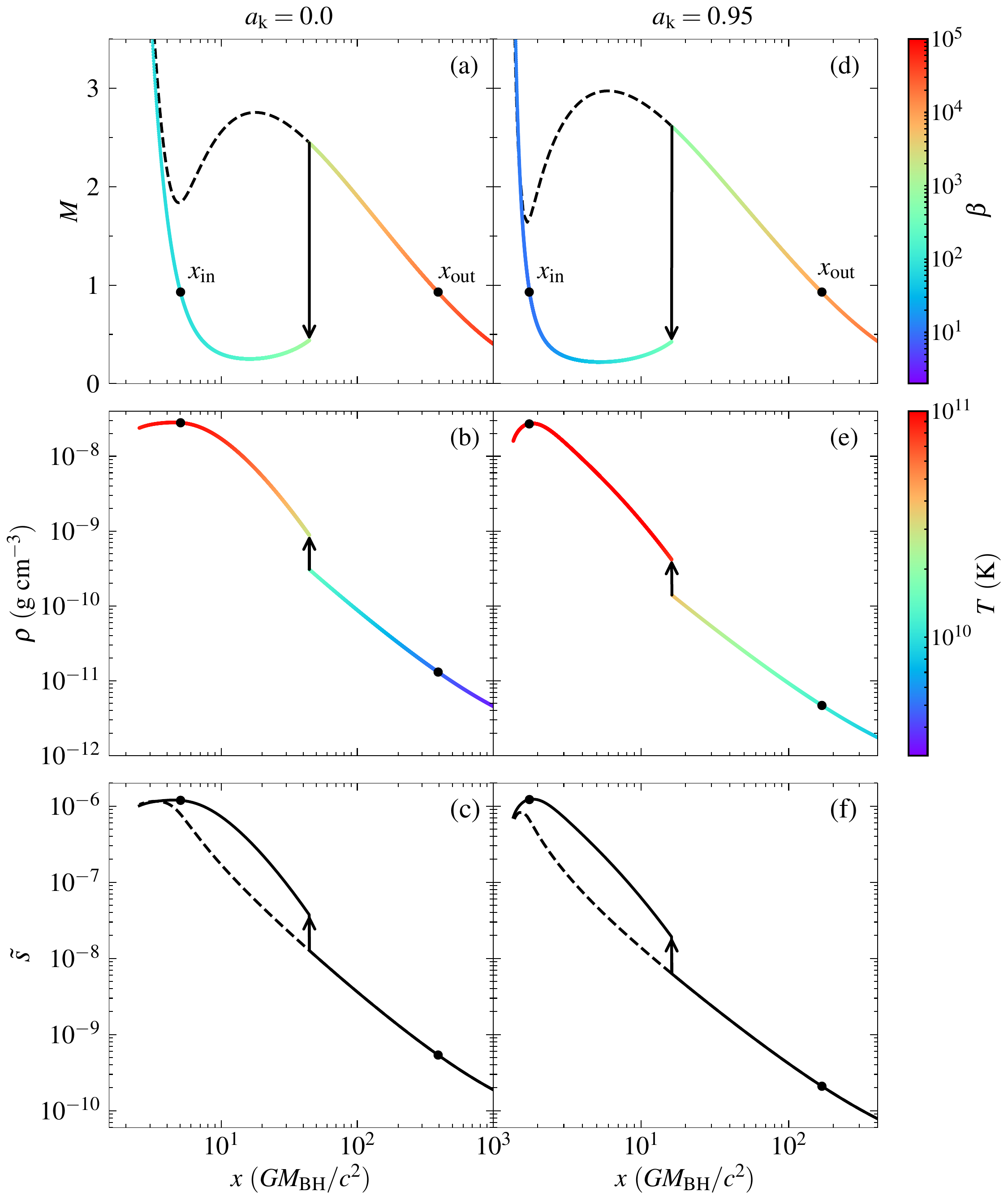}
    \end{center}
    \caption{Plot of shock-mediated global accretion solution around non-rotating ($a_{\rm k} = 0.0$; left panels) and rapidly rotating ($a_{\rm k} = 0.95$; right panels) black holes. Panels (a) and (d) present the variation of the Mach number ($M$), while panels (b) and (e) display the corresponding density profiles ($\rho$) as functions of the radial coordinate ($x$). The colour bars in the upper and lower panels indicate the distribution of plasma-$\beta$ and temperature ($T$) of the accretion flow, respectively. In panels (a) and (d), the colored curves denote shocked solutions, whereas the dashed curves represent shock free solutions. The entropy densities ($\tilde{s}$) corresponding to solutions in panels (a) and (d) are depicted in panels (c) and (f). The solutions are obtained for $l = 1.0$, $p = 0.1$, and $\alpha_{\rm B} = 0.02$. For the non-rotating case (panels a, b, c), the input parameters at the outer edge ($x_{\rm edge} = 1000$) are $\mathcal{E}_{\rm edge} = 0.00158$, $\lambda_{\rm edge} = 4.818$, $\beta_{\rm edge} = 5 \times 10^{4}$, and $\dot{m}_{\rm edge} = 0.1$. For the rotating case (panels d, e, f), the inflow is injected from $x_{\rm edge} = 400$ with $\mathcal{E}_{\rm edge} = 0.00375$, $\lambda_{\rm edge} = 3.092$, $\beta_{\rm edge} = 2 \times 10^{4}$, and $\dot{m}_{\rm edge} = 0.01$. Filled circles mark the inner ($x_{\rm in}$) and outer ($x_{\rm out}$) critical points, while the vertical arrows denote the locations of the shock transitions. See text for the details.
    }
    \label{fig07}
\end{figure}

In Fig. \ref{fig07}, we present representative examples of an accretion solution that undergoes a shock transition in the presence of wind around weakly rotating ($a_{\rm k} \rightarrow 0$) and rapidly rotating ($a_{\rm k} \rightarrow 0.95$) black holes. For this, in Fig. \ref{fig07}a, we choose $l = 1.0$, $p = 0.1$, $\alpha_{\rm B} = 0.02$, $a_{\rm k} = 0$, and inject matter from the outer edge of the disk $x_{\rm edge}=1000$ with $\mathcal{E}_{\rm edge} = 0.00158$, $\lambda_{\rm edge} = 4.818$, $\beta_{\rm edge} = 5 \times 10^{4}$, and $\dot{m}_{\rm edge} = 0.1$. In Fig. \ref{fig07}a, we show the variation of Mach number ($M$) with radial coordinate $ (x)$. As the flow moves inward from $x_{\rm edge}$, it's radial velocity gradually increases and becomes supersonic after smoothly crossing the outer critical point at $x_{\rm out} = 392.84$. The supersonic flow, instead of continuing to move smoothly inward (dashed curve), undergoes a discontinuous transition at $x_{\rm s} = 44.40$ (vertical arrow) in the form of a shock, becoming subsonic. Subsequently, the subsonic flow again gains its velocity and crosses the inner critical point at $x_{\rm in} = 5.03$, turning supersonic before finally plunging into the black hole. The variation of the plasma-$\beta$ parameter during accretion is indicated in Fig. \ref{fig07}a by the colors, which reveal a noticeable drop in plasma-$\beta$ ($\beta_{-}/\beta_{+}=1.49$) across the shock. Because of the shock transition, post-shock matter decelerates and accumulates, leading to an increase in density. Furthermore, as the kinetic energy of the upstream flow is converted into the thermal energy of the downstream flow, the post-shock temperature rises significantly. Since a sharp rise in density and temperature occurs across the shock, the post-shock flow manifests as a puffed-up, torus-like post-shock corona (PSC), which acts as a reservoir of hot electrons. These electrons efficiently inverse Comptonize the soft photons from the pre-shock region, giving rise to the high-energy radiation commonly observed in black hole X-ray binaries \cite[]{Chakrabarti-Titarchuk1995,Iyer-etal-2015,Nandi-etal-2018}. In Fig. \ref{fig07}b, we show the variation of density ($\rho$) corresponding to the accretion solution presented in Fig. \ref{fig07}a, while the temperature ($T$) distribution is indicated by the color, with the color bar on the right denoting the range of $T$. We further calculate the compression ratio ($R$), which quantifies the density increase across the shock front and is defined as the ratio of post-shock to pre-shock density ($R = \Sigma_{+}/\Sigma_{-}$). Similarly, the shock strength ($S$), which determines the temperature jump, is defined as the ratio of the pre-shock Mach number to the post-shock Mach number ($S = M_{-}/M_{+}$). For this shock-induced global accretion solution, we obtain $R=3.99$ and $S=5.56$, respectively. Following \cite{Das-etal2009,Sadowski-etal-2013}, we calculate the entropy density $[\tilde{s} \propto \frac{\rho}{\gamma - 1} \log (P/\rho^\gamma)]$ for both shocked and shock-free accretion solutions, as shown in Fig. \ref{fig07}c. The entropy profile corresponding to the shocked solution (solid curve) exhibits a sharp rise at the shock location ($x_{\rm s}$), clearly indicating a discontinuous increase in entropy across the shock front. This demonstrates that the entropy of the shocked accretion solution is higher than that of the shock-free (dashed curve) solution.

In the right panels of Fig. \ref{fig07}, we depict the shock-induced global accretion solution around rotating black hole with spin parameter $a_{\rm k} = 0.95$. For rapidly spinning black holes, the accretion flow undergoes a shock transition only when its angular momentum is relatively small \citep{Sen-etal-2022}.  The reduced angular momentum diminishes the centrifugal barrier, causing shocks to occur closer to the black hole. Consequently, we adopt a smaller outer edge for the high spin case. Here, we fix $l = 1.0$, $p = 0.1$, $\alpha_{\rm B} = 0.02$, and set $x_{\rm edge} = 400$, with other input parameters as $\mathcal{E}_{\rm edge} = 0.00375$, $\lambda_{\rm edge} = 3.092$, $\beta_{\rm edge} = 2\times 10^{4}$, and $\dot{m}_{\rm edge} = 0.01$. In Fig. \ref{fig07}d, the subsonic flow from the outer edge first passes through the outer critical point at $x_{\rm out} = 167.80$, becoming supersonic, and then experiences a shock transition at $x_{\rm s} = 16.11$, turning subsonic. Finally, it crosses the inner critical point at $x_{\rm in} = 1.76$ and becomes supersonic before plunging into the black hole. In Fig.~\ref{fig07}e, we display the density profile ($\rho$) corresponding to the accretion solution shown in Fig.~\ref{fig07}d, while the temperature ($T$) is represented through a color map, with the color bar on the right indicating the temperature range. For this solution, the corresponding compression ratio and shock strength are obtained as $R = 4.27$ and $S = 6.15$, respectively. Finally, in Fig. \ref{fig07}f, we display the radial profile of the entropy density ($\tilde{s}$) for the solutions corresponding to Fig. \ref{fig07}d. As in Fig. \ref{fig07}c, a marked discontinuity is observed at $x_{\rm s}$, where the entropy rises sharply. This behavior provides conclusive evidence that shock wave generates a state of higher entropy that distinctly separates this solution from the shock-free case.
 
\begin{figure}
    \begin{center}
        \includegraphics[width = \columnwidth]{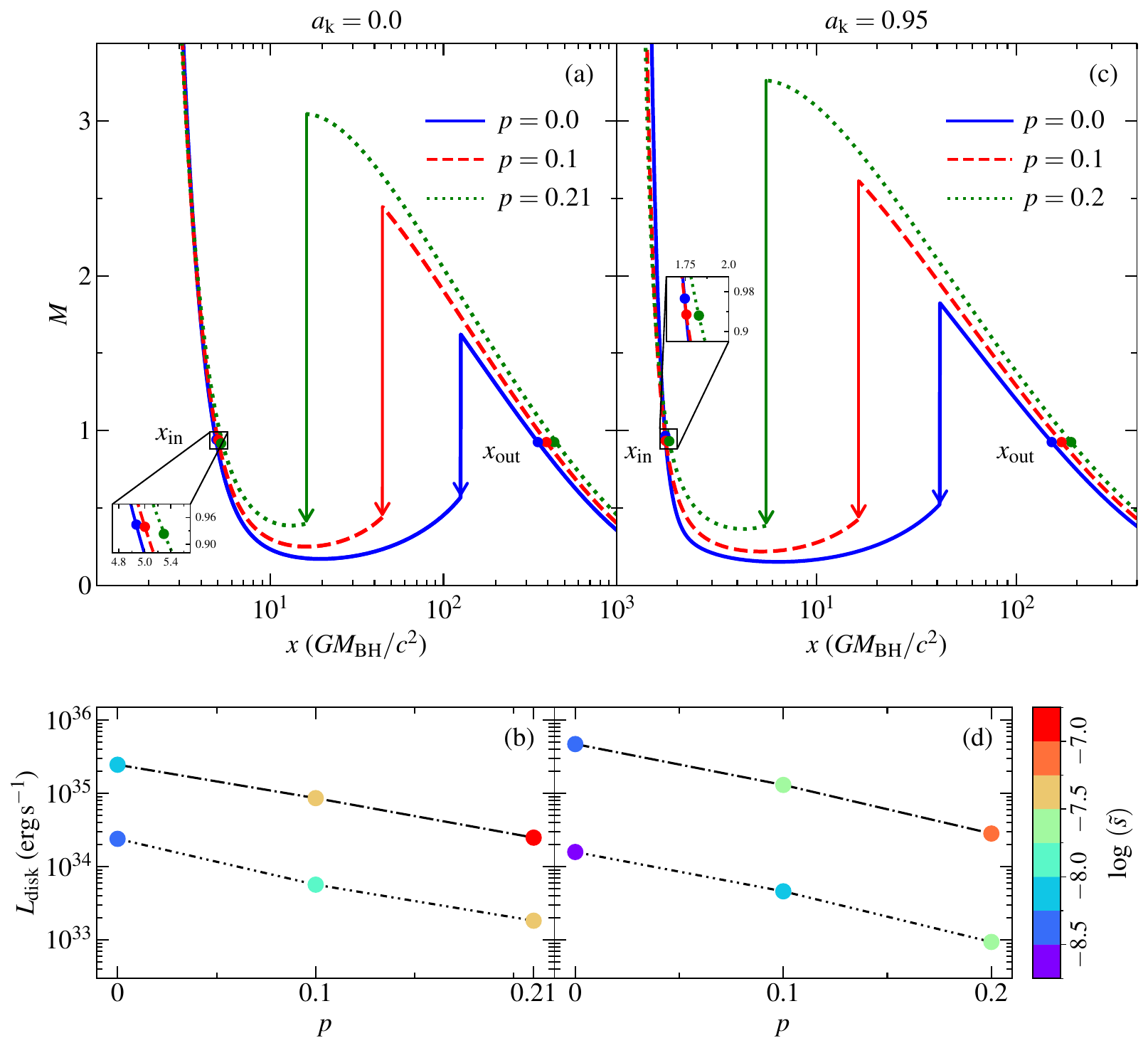}
    \end{center}
    \caption{Modification of shock-induced accretion solutions in the presence of wind. Here, we fix $l = 1.0$ and $\alpha_{\rm B} = 0.02$. In panel (a), a non-rotating ($a_{\rm k} = 0.0$) black hole is considered, and the outer boundary is set at $x_{\rm edge} = 1000$ with input parameters $\mathcal{E}_{\rm edge} = 0.00158$, $\lambda_{\rm edge} = 4.818$, $\beta_{\rm edge} = 5 \times 10^{4}$, and $\dot{m}_{\rm edge} = 0.1$. The solid (blue), dashed (red), and dotted (green) curves are plotted for wind parameters $p = 0$, $0.1$, and $0.21$, respectively. In panel (c), a rapidly rotating black hole with $a_{\rm k} = 0.95$ is chosen, and matter is injected from $x_{\rm edge} = 400$ with $\mathcal{E}_{\rm edge} = 0.00375$, $\lambda_{\rm edge} = 3.092$, $\beta_{\rm edge} = 2 \times 10^{4}$, and $\dot{m}_{\rm edge} = 0.01$. The solid (blue), dashed (red), and dotted (green) curves denote results for $p = 0$, $0.1$, and $0.2$, respectively. In both panels, vertical arrows indicate shock transition radii and filled circles denote the critical points. In panel (b) and (d), variation of disk luminosity ($L_{\rm disk}$) with $p$ is shown. Filled circles joined with dot-dashed line denote results corresponding to the shocked solutions presented in panel (a) and (c), respectively, whereas filled circles joined with dot-dot-dahsed lines are for shock-free solutions. Color code denotes the entropy ($\tilde{s}$) computed immediately after the shock transition. See text for the details.
    }
    \label{fig08}
\end{figure}

We continue to examine the effect of wind on the shock transition in accretion flows. For this analysis, we choose $l = 1.0$ and $\alpha_{\rm B} = 0.02$. The results are presented in Fig. \ref{fig08}, where the left and right panels correspond to black hole spins of $a_{\rm k} = 0.0$ and $0.95$, respectively. In Fig. \ref{fig08}a, the outer boundary is set at $x_{\rm edge} = 1000$, with flow parameters $\mathcal{E}_{\rm edge} = 0.00158$, $\lambda_{\rm edge} = 4.818$, $\beta_{\rm edge} = 5 \times 10^{4}$, and $\dot{m}_{\rm edge} = 0.1$. The results presented using solid (blue), dashed (red), and dotted (green) curves correspond to wind parameters $p = 0.0$, $0.1$, and $0.21$, respectively, while the filled circles denote the locations of the inner and outer critical points. In the absence of wind ($p = 0$), the flow undergoes a shock transition at $ x_{\rm s} = 125.67$. As the wind strength increases to $p = 0.1$ and $p = 0.21$, the shock front shifts inward to $x_{\rm s} = 44.40$ and $x_{\rm s} = 15.94$, respectively. For $p>0.21$, the shock ceases to exist as the Rankine–Hugoniot conditions (RHCs) are no longer satisfied. We subsequently calculate the disk luminosity ($L_{\rm disk}$) for the shocked accretion solutions shown in Fig.~\ref{fig08}a, as well as for the corresponding shock-free cases, and plot the variation of $L_{\rm disk}$ as a function of $p$ in Fig.~\ref{fig08}b. The filled circles connected by dot-dashed and dot-dot-dashed lines represent the shocked and shock-free solutions, respectively. The colour represents the entropy ($\tilde{s}$) evaluated immediately after the shock transition. It is evident that the shocked accretion solutions yield higher $L_{\rm disk}$ and $\tilde{s}$ values compared to the shock-free solutions. Similarly, in the right panel, the outer boundary is fixed at $x_{\rm edge} = 400$, with parameters $\mathcal{E}_{\rm edge} = 0.00375$, $\lambda_{\rm edge} = 3.092$, $\beta_{\rm edge} = 2 \times 10^{4}$, and $\dot{m}_{\rm edge} = 0.01$. In Fig. \ref{fig08}c, solid (blue), dashed (red), and dotted (green) curves denote results obtained for $p = 0.0$, $0.1$, and $0.2$, yielding shock locations at $ x_{\rm s} = 41.26$, $16.11$, and $5.58$, respectively. Note that for $p>0.20$, shock disappears as RHCs become unfavorable. The variation of $L_{\rm disk}$ for shocked and shock-free solutions is shown in Fig.~\ref{fig08}d, similar to Fig.~\ref{fig08}b, exhibiting consistent trends. Overall, these results clearly demonstrate that mass loss through wind significantly affects the structure of the shock-induced global accretion solutions in both non-rotating and rotating black hole. The inward shift of the shock front with increasing $p$ is attributed to a reduction in the post-shock thermal pressure. As more mass is lost through the wind, the density and temperature of the inflow decrease, leading to a weaker pressure support. Consequently, the shock moves closer to the black hole and eventually settles down at smaller radii. The details of model parameters and obtained shock properties are listed in Table \ref{table-1}.

\begin{table*}
	\caption{Details of the model input parameters and shock properties of the shock-induced accretion solutions shown in Fig. \ref{fig08}. Columns $1-2$ list BH spin ($a_{\rm k}$) and wind parameter ($p$). Columns $3-6$ present the flow variables at inner critical points ($x_{\rm in}$), namely $x_{\rm in}$, angular momentum ($\lambda_{\rm in}$), plasma-$\beta_{\rm in}$, and temperature $(T_{\rm in})$. Columns $7-10$ provide the corresponding quantities at outer critical point as $x_{\rm out}$, $\lambda_{\rm out}$, $\beta_{\rm out}$, and $T_{\rm out}$. Finally, columns $11-13$ present the shock location ($x_{\rm s}$), compression ratio ($R$), and shock strength ($S$). See text for the details.}
    \hskip -2.5cm
	\resizebox{2.45\columnwidth}{!}{%
	\begin{tabular}{lcccccccccccc} 	\hline 	\hline
	$a_{\rm k}$ & $p$ & $ x_{\rm in}$ & $ \lambda_{\rm in}$ & $ \beta_{\rm in}$ & $ T_{\rm in}$ & $x_{\rm out}$ & $\lambda_{\rm out}$ & $ \beta_{\rm out}$ & $T_{\rm out}$ & $ x_{\rm s}$ & $R$ & $S$   \\

    & & $ (GM_{\rm BH}/c^2)$ & $ (GM_{\rm BH}/c)$ &  & $(\rm K)$ & $(GM_{\rm BH}/c^2)$ & $(GM_{\rm BH}/c)$ & & (K) & $ (GM_{\rm BH}/c^2)$ & &   \\ \hline
				
	$0.0$ & $0.0$ & $4.90$ & $3.55$ & $109.70$ & $8.41 \times 10^{10}$ & $350.07$ & $3.94$   & $2.19 \times 10^4$ & $5.23 \times 10^9 $ & $125.67$ & $2.44$&$2.85$   \\

    $0.0$ & $0.10$ & $5.03$ & $3.51$ & $77.01$ & $9.30 \times 10^{10}$ & $392.84$ & $4.01$   & $2.07 \times 10^4$ & $4.87 \times 10^9 $ & $44.40$ & $3.99$ & $5.56$   \\

    $0.0$ & $0.21$ & $5.29$ & $3.45$ & $52.82$ & $9.94 \times 10^{10}$ & $445.16$ & $4.08$   & $2.02 \times 10^4$ & $4.52 \times 10^9 $ & $15.94$ & $4.87$ & $7.66$   \\ \hline

    $0.95$ & $0.0$ & $1.75 $ & $2.33 $ & $ 1.12 $ & $3.51 \times 10^{10}$ & $149.76$ & $2.60$ & $9.19 \times 10^3$ & $1.23 \times 10^{10} $ & $41.26$ & $2.86$ & $3.48$ \\

    $0.95$ & $0.10$ & $1.76 $ & $2.30 $ & $10.33  $ & $1.95 \times 10^{11}$ & $167.80$ & $2.63$ & $8.76 \times 10^3$ & $1.14 \times 10^{10} $ & $16.11$ & $4.27$ & $6.15$ \\

    $0.95$&$0.2$&$ 1.83$ & $ 2.26$ & $ 13.86 $ & $2.88 \times 10^{11}$ & $187.29$ &$2.67$   & $8.62\times 10^{3}$ & $1.08 \times 10^{10} $ & $5.58$ & $5.14$ & $8.44$ \\ \hline
			\end{tabular}	
		}
	\label{table-1}
\end{table*}

\begin{figure}
    \begin{center}
        \includegraphics[width = \columnwidth]{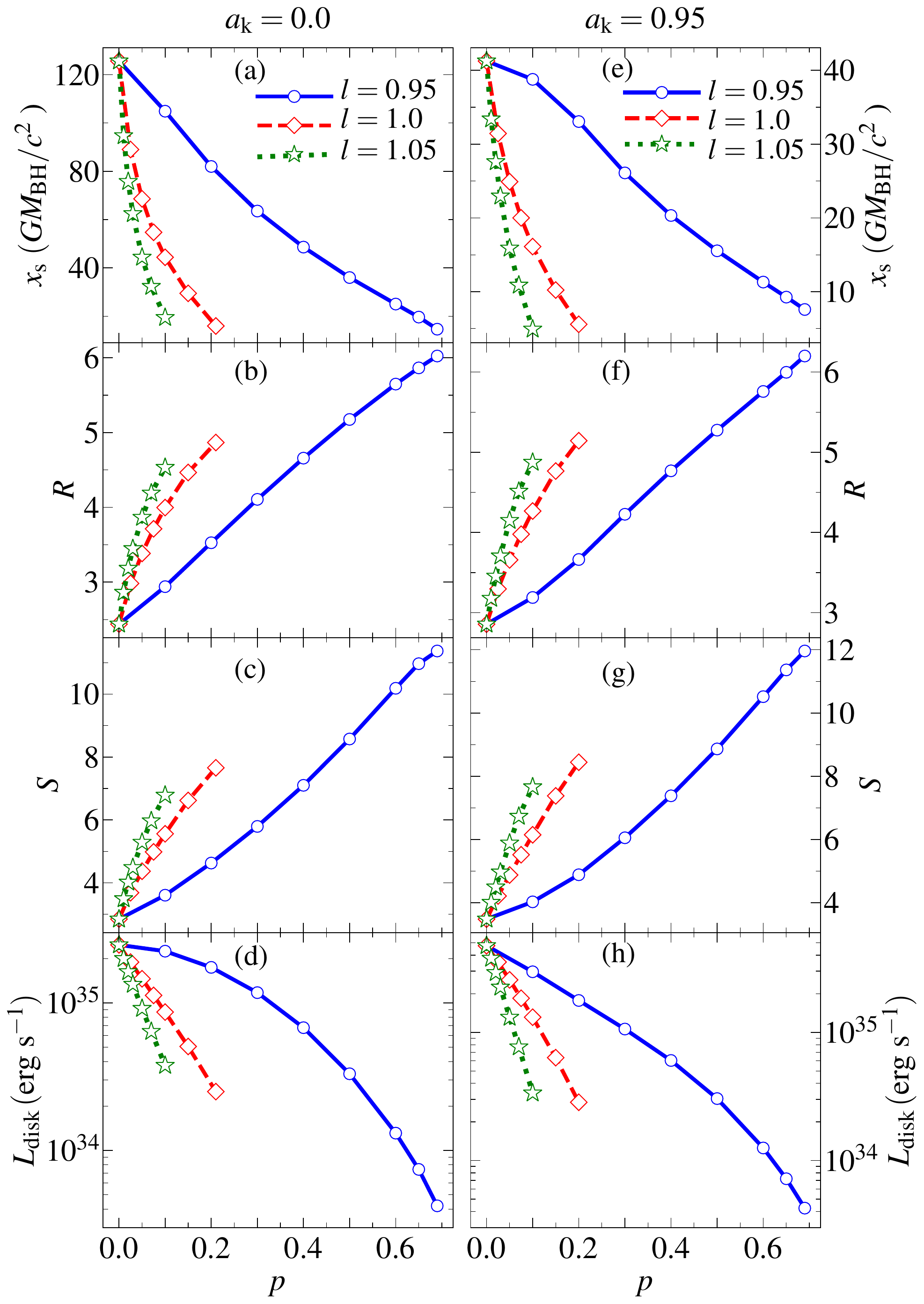}
    \end{center}
    \caption{Variation of (a,e) shock location ($x_{\rm s}$), (b,f) compression ratio ($R$), (c, g) shock strength ($S$) and (d, h) disk luminosity ($L_{\rm disk}$) as a function of wind parameter ($p$) for different values of $l$. The left panels correspond to $a_{\rm k} = 0.0$, where we choose $\alpha_{\rm B} = 0.02$, $x_{\rm edge} = 10^3$, $\mathcal{E}_{\rm edge} = 0.00158$, $\lambda_{\rm edge} = 4.818$, $\beta_{\rm edge} = 5\times 10^4$, and $\dot{m}_{\rm edge} = 0.1$. In the right panels, we choose $a_{\rm k} = 0.95$, $\alpha_{\rm B} = 0.02$ and set $x_{\rm edge} = 400$ with $\mathcal{E}_{\rm edge} = 0.00375$, $\lambda_{\rm edge} = 3.092$, $\beta_{\rm edge} =  2\times 10^{4}$, and $\dot{m}_{\rm edge} = 0.01$. In all panels, circles, diamonds, and asterisks connected with solid (blue), dashed (red), and dotted (green) lines denote results obtained for $l = 0.95$, $1.0$, and $1.05$, respectively. See text for the details.
    }
    \label{fig09}
\end{figure}

Next, we investigate how the shock properties, namely $x_s$, $R$ and $S$, along with the disk luminosity ($L_{\rm disk}$) corresponding to the shocked accretion solution, vary with the wind parameter $p$ for a given outer boundary condition. In doing this, we inject matter from $x_{\rm edge} = 1000$ with $\mathcal{E}_{\rm edge} = 0.00158$, $\lambda_{\rm edge} = 4.818$, $\beta_{\rm edge} = 5\times 10^4$, and $\dot{m}_{\rm edge} = 0.1$, while keeping the viscosity parameter fixed at $\alpha_{\rm B} = 0.02$ for non rotating BH. The results are presented in the left panels of Fig.\ref{fig09}, where the circles, diamonds, and asterisks joined with solid (blue), dashed (red), and dotted (green) lines are for $l = 0.95$, $1.0$, and $1.05$, respectively. Similar to the behavior shown in Fig. \ref{fig08}a, we find that, for all values of $l$, the shock front gradually moves towards the black hole, as $p$ increases (see Fig. \ref{fig09}a). We also notice that for a given $p$, $x_{\rm s}$ decreases with increasing $l$. This trend arises because higher values of $l$ allow the outflow to extract more angular momentum from the accretion flow, thereby reducing the angular momentum of the inflowing matter. As a result, the centrifugal support weakens, causing the shock to settle down closer to the black hole. Further, in Fig. \ref{fig09}b, we show the variation of the compression ratio ($R$) with wind parameter ($p$) for shocked accretion solutions presented in Fig. \ref{fig09}a. As the inflowing matter loses more mass through winds, the shock shifts inward, resulting in stronger compression in the PSC region, and consequently, $R$ increases. Similarly, the increase in $l$ pushes the shock front towards the BH resulting in higher compression ratio. In Fig. \ref{fig09}c, we present the variation of shock strength ($S$) with $p$ for the same shocked accretion solutions and observe that $S$ increases with both $p$ and $l$. Finally, Fig. \ref{fig09}d illustrates the variation of disk luminosity, where it is observed that $L_{\rm disk}$ decreases as $p$ and $l$ increase, consistent with the trends seen in Fig. \ref{fig05} and \ref{fig06}. We extend this study for rotating BHs with $a_{\rm k} = 0.95$ and inject matter from $x_{\rm edge} = 400$ with $\mathcal{E}_{\rm edge} = 0.00375$, $\lambda_{\rm edge} = 3.092$, $\beta_{\rm edge} = 2 \times 10^{4}$, $\dot{m}_{\rm edge} = 0.01$ and $\alpha_{\rm B} = 0.02$. The obtained results are depicted in the right panels of Fig. \ref{fig09}, where circles, diamonds, and asterisks, connected by solid (blue), dashed (red), and dotted (green) lines correspond to $l = 0.95$, $1.0$, and $1.05$, respectively. It is evident from Figs. \ref{fig09}e-h that the overall trends of $x_{\rm s}$, $R$, $S$, and $L_{\rm disk}$ with $p$ and $l$ are qualitatively similar to those obtained for the non-rotating black hole in the left panels. These results clearly indicate that both $p$ and $l$ play a crucial role in regulating the shock properties and disk luminosity.

\begin{figure}
    \begin{center}
        \includegraphics[width = \columnwidth]{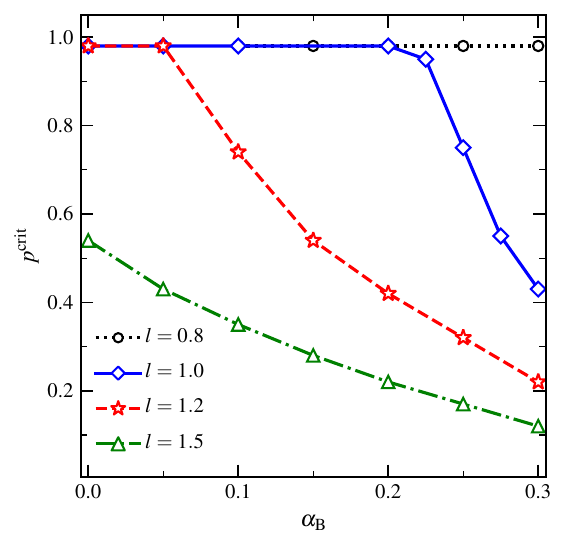}
    \end{center}
    \caption{Variation of the critical wind parameter $p^{\rm crit}$ as a function of $\alpha_{\rm B}$ for different values of $l$. Here, $\beta_{\rm in} = 100$, $\dot{m}_{\rm in} = 0.01$, and $a_{\rm k} = 0.95$ are adopted. See text for the details.
    }
    \label{fig10}
\end{figure}

Finally, we investigate the allowed range of the wind parameter $p$ that admits shock-induced accretion solutions, and examine how its threshold value ($p^{\rm crit}$) varies with $\alpha_{\rm B}$ for different choices of $l$ parameters. We carry out this analysis for a rotating black hole with spin $a_{\rm k}=0.95$, and choose $\beta_{\rm in}=100$ and $\dot{m}_{\rm in}=0.01$ at the inner critical point $x_{\rm in}$. By varying the flow energy ($\mathcal{E}_{\rm in}$) and angular momentum ($\lambda_{\rm in}$) at $x_{\rm in}$, we determine $p^{\rm crit}$ for given values of $\alpha_{\rm B}$ and $l$. The obtained results are presented in Fig.~\ref{fig10}, where circles, diamonds, asterisks, and triangles joined using dotted (black), solid (blue), dashed (red), and dot-dashed (green) lines correspond to results for $l=0.8,\,1.0,\,1.2$, and $1.5$, respectively. We observe that for $l=0.8$, $p^{\rm crit}$ reaches its maximum value ($p^{\rm crit}=1$) irrespective of the viscosity parameter $\alpha_{\rm B}$. In contrast, for $l=1.0$ and $l=1.2$, $p^{\rm crit}$ initially remains insensitive to $\alpha_{\rm B}$ (i.e., $p^{\rm crit}=1$), and subsequently decreases as $\alpha_{\rm B}$ increases. Indeed, the decrease of $p^{\rm crit}$ from its maximum value begins at comparatively lower $\alpha_{\rm B}$ when $l$ is larger. This implies that accretion flows can sustain shocks with $p^{\rm crit}=1$ over a broader range of $\alpha_{\rm B}$ for lower values of $l$. For $l=1.5$, shock-induced accretion solutions exist with $p^{\rm crit}=0.54$ in the weak-viscosity limit ($\alpha_{\rm B} \rightarrow 0$), after which $p^{\rm crit}$ decreases with increasing $\alpha_{\rm B}$. It is worth noting that for $l>1$, the inflowing matter loses part of its angular momentum as it is carried away by the winds for a fixed $p$. Furthermore, as $\alpha_{\rm B}$ increases, outward transport of angular momentum becomes more efficient, leading to an overall reduction of angular momentum at the inner edge of the disk. Consequently, the centrifugal repulsion against gravity weakens, and the possibility of shock formation reduces, resulting in a decrease of $p^{\rm crit}$.

\section{Conclusions}
\label{concl}

In this study, we investigate the effect of mass loss on the dynamics of viscous, magnetized, advective accretion flows around black holes. The accretion flow is assumed to be axisymmetric and primarily threaded by the toroidal component of the magnetic field, with synchrotron emission serving as the dominant radiative cooling mechanism. To avoid the complexities of full general relativistic treatment while retaining essential relativistic effects, we employ an effective potential \citep{Dihingia-etal2018a} that satisfactorily reproduces the space–time geometry around a rotating black hole. Mass loss is incorporated by allowing the accretion rate to vary radially as a power law, $\dot{M}_{\rm a} \propto x^p$, following \citet{Blandford-Begelman1999}, where $0 \leq p < 1$. The fraction of angular momentum extracted by the wind is parameterized according to \citet{Knigge-1999} and \citet{Xie-Yuan-2008}. The main findings of this work are summarized below.

\begin{itemize}

    \item We find that the presence of winds substantially alters the structure of the accretion solutions. Mass loss leads to a significant reduction in the density, temperature, and plasma-$\beta$ parameter of the flow. The rotational velocity of the inflowing matter, however, exhibits a strong dependence on the wind parameter $l$, which determines the specific angular momentum carried away by the winds. For $l < 1$, the flow spins up as mass loss increases, whereas for $l > 1$, the disk rotation slows down with stronger outflows (see Fig. \ref{fig03} and Fig. \ref{fig04}). These findings clearly demonstrate the crucial role of wind in regulating the dynamical structure of accretion disks.

    \item We compute the disk luminosity and find that it exhibits a strong dependence on both the mass and angular momentum carried away by the wind (see Fig. \ref{fig05} and Fig. \ref{fig06}). An increase in mass loss results in a significant reduction of the radiative power, suggesting that disk winds may play a crucial role in producing the observed sub-Eddington luminosities in many accreting black hole systems.

    \item We further extend our analysis to obtain the global transonic accretion solutions containing shocks (see Fig. \ref{fig07}) around both non-rotating and rotating black holes. The results indicate that the presence of wind modifies the shock structure while preserving the general trend that shocked accretion solutions remain more luminous than their shock-free solutions. As the wind parameters $p$ and $l$ increase, the shock front systematically shifts inward, resulting in a more compact post-shock corona (PSC). This behavior arises because mass loss reduces the post-shock thermal pressure and angular momentum loss weakens the centrifugal support, thereby forcing the shock to form closer to the black hole (see Fig. \ref{fig08}). Consequently, the compression ratio ($R$) and shock strength ($S$) increase, while the overall disk luminosity ($L_{\rm disk}$) diminishes. These results collectively demonstrate that both mass and angular momentum loss through winds play a decisive role in regulating the dynamics, structure, and radiative properties of shocked accretion flows (see Fig. \ref{fig09}).

    \item Finally, we identify the critical mass-loss parameter ($p^{\rm crit}$) that allows the existence of steady shock solutions. Our analysis reveals that higher viscous stresses tend to suppress mass loss, while enhanced angular momentum extraction by the wind reduces $p^{\rm crit}$. These correlations emphasize the strong coupling among viscosity, mass outflow, and shock formation in black hole accretion flows (see Fig. \ref{fig10}).

\end{itemize}

It is worth noting that the present study is carried out under several simplifying assumptions, and the adiabatic index ($\gamma$) is held fixed rather than determined self-consistently from local thermodynamic conditions. We consider only the toroidal component of the magnetic field, neglecting the contribution of the poloidal field. Synchrotron radiation is treated as the sole cooling mechanism, while bremsstrahlung and Compton processes have been ignored. Energy loss directly associated with the wind is also not included. While all these physical processes are indeed relevant, their implementation lies beyond the scope of the present work and will be addressed in future studies.

\section*{Data Availability}

The numerical datasets underlying this work are substantial in size and therefore cannot be deposited in a public repository. The data are, however, available upon reasonable request.

\section*{Acknowledgments}

Authors thank the anonymous reviewer for constructive comments and useful suggestions that helped to improve the quality of the manuscript. Authors acknowledge the support from the Department of Physics, IIT Guwahati for providing the facilities to complete this work.

\bibliography{arxiv_version.bbl}

\appendix

\section*{Derivation of flow variable gradients}

Using equation (\ref{power2}) in equations (\ref{radialmom}), (\ref{angular}), (\ref{entropy}) and (\ref{magflux}), we get
$$
\mathcal{E}_{\rm v} \frac{dv}{dx} + \mathcal{E}_{\rm cs} \frac{dC_{\rm s}}{dx} + \mathcal{E}_{\lambda} \frac{d\lambda}{dx} + \mathcal{E}_{\beta} \frac{d\beta}{dx} + \mathcal{E}_0 = 0,
\eqno(A1)
$$
$$
l_{\rm v} \frac{dv}{dx} + l_{\rm cs} \frac{dC_{\rm s}}{dx} + l_{\lambda} \frac{d\lambda}{dx} + l_{\beta}\frac{d \beta}{dx} + l_0= 0,
\eqno(A2)
$$
$$
R_{\rm v} \frac{dv}{dx} + R_{\rm cs} \frac{dC_{\rm s}}{dx} + R_{\lambda} \frac{d\lambda}{dx} + R_{\beta}\frac{d \beta}{dx} + R_0= 0,
\eqno(A3)
$$
$$
b_{\rm v} \frac{dv}{dx} + b_{\rm cs} \frac{dC_{\rm s}}{dx} + b_{\lambda} \frac{d\lambda}{dx} + b_{\beta}\frac{d \beta}{dx} + b_0= 0.
\eqno(A4)
$$

The coefficients of equations ($A1-A4$) are expressed in the form which are given by,
\begin{align*}
& \mathcal{E}_{v} = \frac{\gamma v^2 - C^2_{\rm s}}{\gamma v}, \quad \mathcal{E}_{cs} = \frac{C_{\rm s}}{\gamma}, \quad \mathcal{E}_{\lambda} = \frac{C^2_{\rm s}}{2 \gamma \mathcal{F}} \frac{\partial \mathcal{F}}{\partial \lambda}\Big|_{x}, \quad \mathcal{E}_{\beta} = 0, \quad
\mathcal{E}_{0} = \frac{p C^2_{\rm s}}{\gamma x} + \frac{C^2_{\rm s}}{2 \gamma \mathcal{F}} \frac{\partial \mathcal{F}}{\partial x}\Big|_{\lambda} - \frac{3 C^2_{\rm s}}{2 \gamma x}  - \frac{C^2_{\rm s}} { \gamma x} + \frac{d \Psi^{\rm eff}}{dx} + \frac{2 C^2_{\rm s}}{\gamma (1 + \beta) x},\\
&l_v = \alpha_{\rm B} x \Big(1 - \frac{g C^2_{\rm s}}{\gamma v^2}\Big), \quad g = \frac{I_{n+1}}{I_n}, \quad l_{\rm cs} = \frac{2 \alpha_{\rm B} x g C_{\rm s}}{\gamma v}, \quad l_{\lambda} = -1, \quad l_{\beta} = 0, \quad l_0 = \frac{\alpha_{\rm B}}{\gamma v}(g C^2_{\rm s} + \gamma v^2)(1 + p ) + \frac{\lambda p}{x}(l - 1),\\
& R_v = \frac{C^2_{\rm s}}{\gamma}\frac{\beta}{1+ \beta}, \quad R_{\rm cs}=\frac{\gamma+1}{\gamma-1}\frac{C^2_{\rm s} v}{\gamma}\frac{\beta}{1+ \beta}, \quad R_{\lambda} = -\Big(\frac{C^2_{\rm s} v}{\gamma}\frac{\beta}{1+ \beta}\frac{1}{2 \mathcal{F}}\frac{\partial \mathcal{F}}{\partial \lambda}\Big|_{x}  + \frac{2 \alpha_{B} I_n}{\gamma} (g C^2_{\rm s} + \gamma v^2) x \frac{\partial \Omega}{\partial \lambda}\Big|_{x}\Big), \\
& R_\beta = \frac{C^2_{\rm s} v}{\gamma (\gamma-1)(1 + \beta)^2}, \quad  R_0 = \frac{C^2_{\rm s} v}{\gamma}\frac{\beta}{1+ \beta}\Big(  \frac{5}{2 x} -\frac{1}{2 \mathcal{F}}\frac{\partial \mathcal{F}}{\partial x}\Big|_{\lambda} - \frac{p}{x} \Big) - \frac{S' C^5_{\rm s}}{v}\sqrt{\frac{\mathcal{F}}{x^5 }}\frac{\beta ^2}{(1 + \beta)^3} - \frac{2 \alpha_{B} I_{n}}{\gamma}(g C^2_{\rm s} + \gamma v^2) x \frac{\partial \Omega}{\partial x}\Big|_{\lambda},\\
& b_v = 1/v, \quad b_{\rm cs} = 3/C_{\rm s}, \quad b_{\lambda} = \frac{-1}{2 \mathcal{F}}\frac{\partial \mathcal{F}}{\partial \lambda}\Big|_{x}, \quad b_{\beta} = -1/(1+ \beta), \quad b_0 = \frac{(p  +2\zeta-1)}{x}  + \frac{3}{2x} -\frac{1}{2 \mathcal{F}}\frac{\partial \mathcal{F}}{\partial x}\Big|_{\lambda}.
\end{align*}

The coefficients mentioned in equations (\ref{grad_a}), (\ref{grad_l}) and (\ref{grad_b})
 are given by,
\begin{align*}
& C_{\rm s0} = -\frac{\mathcal{E}_{\lambda} l_0 + \mathcal{E}_0}{\mathcal{E}_{\rm cs} + \mathcal{E}_{\lambda} l_{\rm cs}},\quad C_{\rm sv} = -\frac{\mathcal{E}_{\lambda} l_v + \mathcal{E}_v}{\mathcal{E}_{\rm cs} + \mathcal{E}_{\lambda} l_{\rm cs}}, \quad
\lambda_{0} = C_{\rm s0} l_{\rm cs} +l_0, \quad \lambda_{v} = C_{\rm sv} \, l_{\rm ca} + l_v,\\
& \beta_{0} = - \frac{b_0 + b_{\lambda} \, \lambda_{0} + b_{\rm cs} \, C_{\rm s0}}{b_{\beta}}, \quad
\beta_{v} = -\frac{b_v + b_{\lambda} \, \lambda_{v}+ b_{\rm cs} \, C_{\rm sv} }{b_{\beta}}.
\end{align*}

By utilizing the aforementioned coefficients, the numerator ($\mathcal{N}$) and denominator ($\mathcal{D}$) of equation (\ref{grad_v}) are expressed as,
\begin{align*}
N(x, v, C_{\rm s}, \lambda, \beta)  &= - (R_0 + R_{\rm cs} \,C_{\rm s0} + R_{\lambda} \lambda_{0} + R_{\beta}\, \beta_{0}),\\
D(x, v, C_{\rm s}, \lambda, \beta) &=  (R_v + R_{\rm cs} \,C_{\rm sv} + R_{\lambda} \, \lambda_{\rm v} + R_{\beta} \, \beta_{\rm v}).
\end{align*}

\end{document}